\documentclass[review]{elsarticle}
\graphicspath{ {./figures/} }
\usepackage{hyperref}
\usepackage{float}
\usepackage{verbatim} 
\usepackage{apalike}

\usepackage{amsfonts}
\usepackage{amssymb}
\usepackage{amsmath}
\usepackage{booktabs}
\usepackage{makecell}
\usepackage{tabularx}
\usepackage{dsfont}
\usepackage{graphicx} 
\restylefloat{figure}
\restylefloat{table}

\journal{Expert Systems with Applications}

\biboptions{authoryear}

\begin{document}
\begin{frontmatter}

\title{Video2Music: Suitable Music Generation from Videos using an Affective Multimodal Transformer model}

\author[label1]{Jaeyong Kang \corref{cor1}}
\ead{kjysmu@gmail.com}

\author[label1]{Soujanya Poria}
\ead{sporia@sutd.edu.sg}

\author[label1]{Dorien Herremans}
\ead{dorien\_herremans@sutd.edu.sg}

\cortext[cor1]{Corresponding author.}
\address[label1]{Singapore University of Technology and Design, 8 Somapah Rd, Singapore 487372}

\begin{abstract}
Numerous studies in the field of music generation have demonstrated impressive performance, yet virtually no models are able to directly generate music to match accompanying videos. In this work, we develop a generative music AI framework, Video2Music, that can match a provided video. We first curated a unique collection of music videos. Then, we analysed the music videos to obtain semantic, scene offset, motion, and emotion features. These distinct features are then employed as guiding input to our music generation model. We transcribe the audio files into MIDI and chords, and extract features such as note density and loudness. This results in a rich multimodal dataset, called MuVi-Sync, on which we train a novel Affective Multimodal Transformer (AMT) model to generate music given a video. This model includes a novel mechanism to enforce affective similarity between video and music. Finally, post-processing is performed based on a biGRU-based regression model to estimate note density and loudness based on the video features. This ensures a dynamic rendering of the generated chords with varying rhythm and volume. 
In a thorough experiment, we show that our proposed framework can generate music that matches the video content in terms of emotion. The musical quality, along with the quality of music-video matching is confirmed in a user study. The proposed AMT model, along with the new MuVi-Sync dataset, presents a promising step for the new task of music generation for videos.
\end{abstract}

\begin{keyword}
Generative AI \sep Music Generation \sep Transformer \sep Multimodal \sep Affective Computing \sep Music Video Matching
\end{keyword}

\end{frontmatter}

\section{Introduction}
\label{sec:1}
In today's digital era, social media platforms such as YouTube have revolutionized the way that videos are consumed and shared. These platforms have given rise to a new form of entertainment, where captivating visuals are often complemented by carefully curated background music. While the advancements in mobile device technology have made it easier than ever to capture high-quality videos, the challenge of finding suitable background music that perfectly aligns with the video content remains a daunting task, and such music is often subject to copyright. In this work, we aim to provide a solution for this by developing a framework, called Video2Music, for music generation to match video. 

The integration of suitable background music in videos is crucial in elevating the overall viewer experience as well as eliciting the desired emotional response. A well-chosen soundtrack can enhance the storytelling, reinforce the mood, and intensify the impact of the visual narrative~\citep{littlefield1990unheard}. However, the process of hand-selecting music tracks that synchronize perfectly with the visual elements of a video is far from trivial. It requires a deep understanding of musical composition, genre, tempo, and the ability to discern the intricate nuances and dynamics of both the video and the accompanying music. In addition, since this music is often pre-composed, its mood and tempo do not dynamically adapt to the video. 

The issue of copyright further compounds the complexity of this endeavor. The availability and licensing restrictions associated with commercially produced music tracks limit the choices available to video creators, often compromising the level of alignment and cohesiveness between the visuals and the music. Although there has been some recent work on music matching/recommendation for video~\citep{thao2023emomv}, solely recommending existing music does not overcome these issues. 

There are very few studies that tackle the task of music generation for video. Some of the pioneering attempts include models like V2Meow~\citep{su2023v2meow} and Controllable Music Transformer (CMT)~\citep{di2021video}, but these are few and far between, underlining the vast potential that lies in this under-explored territory. The recent V2Meow model~\citep{su2023v2meow} synthesizes waveforms directly from video. Music generation models have typically been MIDI-based~\citep{herremans2017functional}, as this offers a finer-grained control and offers composers the opportunity to use the generated MIDI in their Digital Audio Workstations (DAWs). Other work, such as~\citet{di2021video}'s CMT offers a promising first step in MIDI generation for music. Their model does not use a joint music-video dataset, but instead first defines the relationship between music and video based on three characteristics: timing, motion speed, and motion saliency. 

A number of limitations have hindered the development of existing music generation models for video. First and foremost, the scarcity of comprehensive and diverse datasets that incorporate both audio MIDI as well as synced video has hampered the advancement of this field. Furthermore, while a handful of music generation models for video do exist, they remain relatively scarce due to the challenge of effectively synchronizing music with the visual dynamics of videos. 

\begin{figure}[t!] 
        \centering 
        \includegraphics[width=12 cm]{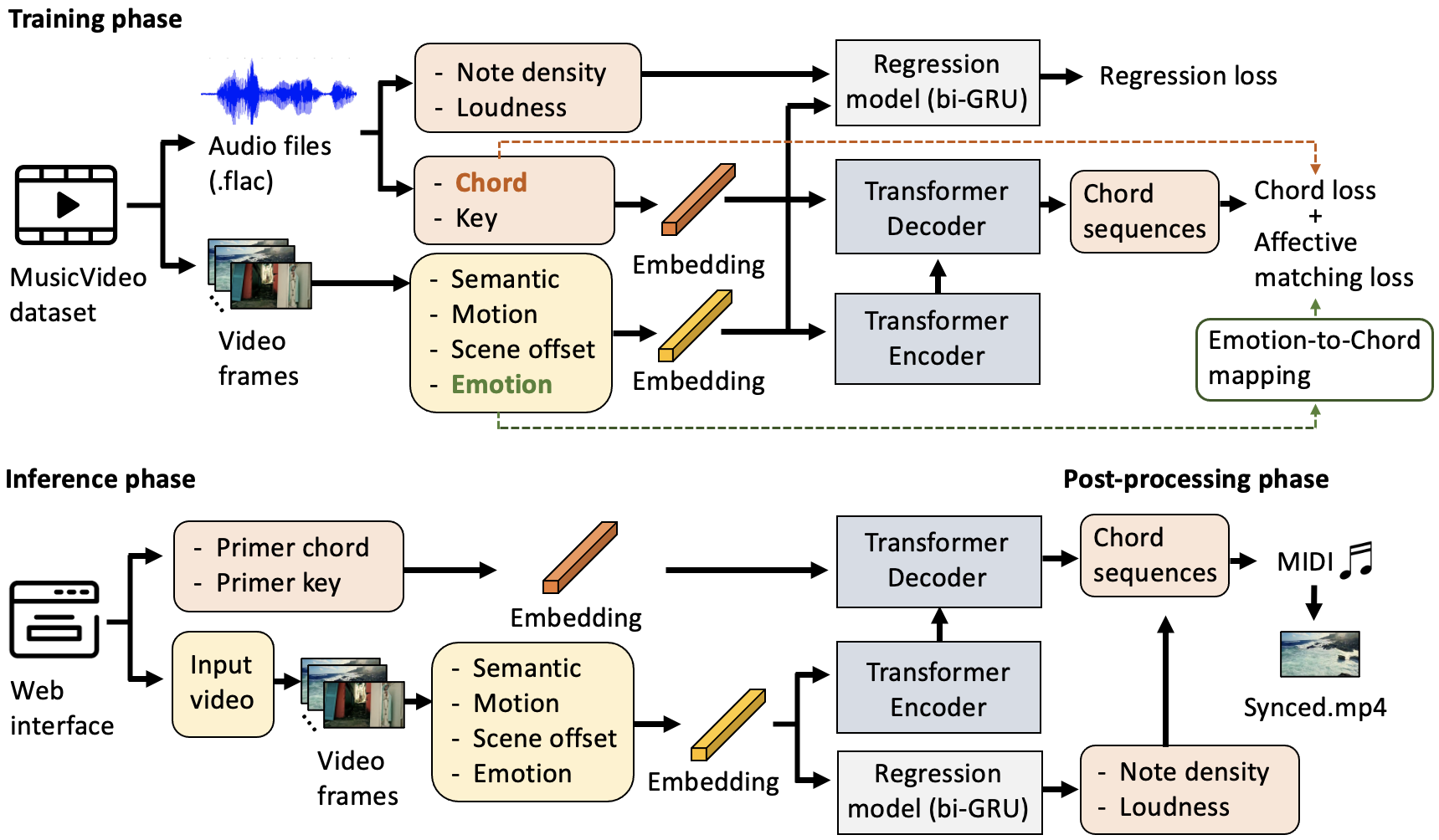} 
        \caption{Overview of our proposed Video2Music. In the training phase, we extract features from the audio file as well as the video frames and subsequently train the transformer model to predict chord sequences given video. We implemented two losses (chord loss and affective matching loss) to train the model. In the inference phase, the uploaded video and primer chords and key from the user are fed into the trained model to generate chord sequences. In the post-processing phase, we estimate note density and loudness from the input video and use them to synthesize the matching MIDI.}
        \label{fig:overview} 
\end{figure}

To address these limitations, we propose a novel AI-powered multimodal music generation framework called Video2Music. This framework uniquely uses video features as conditioning input to generate matching music using a Transformer architecture. By employing cutting-edge technology, our system aims to provide video creators with a seamless and efficient solution for generating tailor-made background music. The overview of our Video2Music framework is shown in Figure~\ref{fig:overview}.

To overcome the scarcity of existing models and the underlying data gap, we introduce MuVi-Sync, a novel dataset comprising popular music videos with a large collection of extracted video and music features.  We opted to work with music videos, as these have been specifically designed with a focus on synchronization between music and video. From the video tracks, we extract semantic, scene offset, motion, and emotion features from these videos. These features serve as essential guidance and conditioning for our music generation model. From the music audio tracks, we transcribe the chords (used for training the generative model) as well as a MIDI file. From these MIDI files, we extract the note density and loudness features. These features are used to construct a biGRU-based regression model for post-processing, which is able to estimate note density and loudness from video features. This mechanism allows for the generation of music with varying rhythms and volume levels. 

The core of our proposed Video2Music framework is a novel Affective Multimodal Transformer (AMT) model, which generates chords given a video. This model consists of two fundamental components: an encoder, which takes the extracted input features from the video as a conditioning factor, and a decoder which takes input features associated with chords and keys extracted from the audio during training as well as conditioning from the decoder, and learns to predict new chords during inference. It is essential to underscore the distinctive features that set our AMT model apart from previous research that uses Multimodal Transformers. Multimodal Transformers have previously been used for a variety of tasks that involve understanding and generating relationships between different modalities of data, such as: Image captioning~\citep{yu2019multimodal, li2019entangled}, Visual question answering~\citep{khan2020mmft,gao2023mist}, and Video summarization~\citep{zhao2022hierarchical,narasimhan2021clip,zhu2023topic}. Notably, our model pioneers the application of Multimodal Transformers to the specific task of music generation for video, addressing the challenge of limited training data in this domain. Our AMT model not only captures intricate relationships between video and music data but also incorporates a novel mechanism for enforcing affective similarity, ensuring a more nuanced and emotionally resonant music-video correspondence. We have set up an extensive experiment, including an objective experiment, as well as a subjective listening study, which shows that our proposed Video2Music framework is able to successfully generate music that matches video, with a quality that outperforms the baseline models.

Our approach is innovative in several ways: 1) Affective Matching: Our AMT model not only captures music-video relationships but also incorporates a novel mechanism for enforcing affective similarity. This ensures that the generated music evokes emotions that resonate with the video's mood and content. 2) Comprehensive Transcribed Dataset: We introduce the MuVi-Sync dataset, an openly available resource containing video and music features extracted from music videos. This rich dataset provides invaluable training data for our model. 3) Expressive Music Generation: Video2Music generates chords, allowing for further customization of expressivity during post-processing. During the latter, a bi-GRU model estimates the note density and loudness based on video features, enabling generation of music with dynamic rhythms and volume levels.

In sum, our music generation system represents a pioneering approach to tackle the novel task of music generation for video. 

Our contributions are as follows:

\begin{enumerate}

    \item We developed one of the first generative music models that match music to a given video, by steering emotional alignment. 
    
	\item We collected an openly available music video dataset, called MuVi-Sync, that consists of video features (scene offset, emotion, motion, and semantic) and music features (chord, key, loudness, and note density).
 
 	\item Our proposed music generation framework utilizes cutting-edge technology: 1) Transformer model for generating chord sequences that match video, and 2) a bi-GRU model for estimating note density and loudness which are used in a post-processing stage. The source code of our proposed framework and the trained models are available online\footnote{\url{https://github.com/AMAAI-Lab/Video2Music}}. 
 
	\item  We conducted extensive experiments and showed that our proposed model outperforms baseline models in terms of music-video correspondence as well as chord prediction accuracy.
\end{enumerate}

In the rest of the paper, we first present the related literature in Section~\ref{sec:2}. This is followed by a description of how we created the new dataset (Section~\ref{sec:3}), after which we present the details of our proposed Video2Music framework in Section~\ref{sec:4}. Finally, we describe our experimental setup (Section~\ref{sec:5}) and its results together with a discussion (Section~\ref{sec:6}). Finally, Section~\ref{sec:7} offers conclusions from this work.

\section{Related Work}
\label{sec:2}
We will provide a brief overview of existing Transformer-based music generation systems, followed by a description of music generation for video. In this section, we do not aim to give an exhaustive overview of music generation systems, for this, the user is referred to~\citet{herremans2017functional}, \citet{civit2022systematic}, and~\citet{briot2020deep}. 

\subsection{Transformer-based Music Generation}
\label{sec:2.1}
Patterns and long-term structure are key features of music~\citep{herremans2017morpheus, herremans2015generating}. Hence, a few years ago, recurrent neural network architectures were welcomed for music generation~\citep{chuan2018modeling, hadjeres2020anticipation, goel2014polyphonic, sturm2019machine}. Up until then, these models typically surpassed other architectures in terms of long-term structure. Other types of models, however, may have their own strengths, e.g. VAEs are known for feature disentanglement~\citep{tan2020music, guo2020variational}, hybrid optimization approaches can constrain patterns \citep{herremans2013composing, herremans2017morpheus}, and embedding methods such as word2vec are known for learning representations~\citep{chuan2020context, huang2016chordripple}. 

In recent years, the Transformer architecture, introduced by~\citet{vaswani2017attention}, has emerged as a dominant force in temporal sequence processing, surpassing traditional Recurrent Neural Networks (RNNs) in various domains. This shift towards Transformers is not confined to the realm of natural language processing; it extends to diverse fields such as computer vision~\citep{arnab2021vivit, ding2023sw}, audio processing~\citep{gong2021ast}, and even reinforcement learning~\citep{chen2021decision}. The self-attention mechanism inherent in Transformers allows them to capture long-range dependencies more effectively, contributing to their success in handling sequential data across different domains. The use of Transformers has also become a trend in the field of music generation, with numerous approaches exploring the potential of Transformers as described in what follows.

\citet{huang2018music} proposed a Music Transformer to generate Chorales as well as classical piano pieces, of length 2,000 tokens. To the best of our knowledge, it is the first Transformer-based model developed to generate music. 

Other models soon followed, for instance, \citet{payne2019musenet} presented MuseNet, a GPT-2-based Sparse Transformer model, that can generate music pieces that are up to 4 minutes long with 10 different instruments and various styles. The Sparse Transformer does not use relative attention, but instead implements full attention over a total of 4,096 tokens. This makes it is better suited for capturing long-term structure. Both MuseNet and Music Transformer use decoder-only Transformers. In both of these models, the teacher-forcing algorithm during training. 

Both of these systems also use a cross-entropy loss, which is not a real measure of musical quality. \citet{zhang2020learning} attempted to solve this issue by proposing an Adversarial Transformer that produces high quality classical guitar music. In the adversarial Transformer model, the self-attention architecture is combined with generative adversarial learning, whereby the generator is a decoder-only Transformer, and the discriminator is an encoder-only Transformer. In addition, adversarial objectives are used as a strong regularization for enforcing the Transformer to focus on learning the local and global structures. 

The success of Transformers for music generation is confirmed by the many models that have followed in the subsequent years. For instance, \citet{wu2020jazz} presented a jazz Transformer for the task of generating monophonic jazz solos, which is based on the Transformer-XL model. Other adversarial models include that of \citet{muhamed2021symbolic}, who developed a model for piano music generation based on adversarial training of a Transformer model. As a generator, they used Transformer-XL and as a discriminator, they used BERT to extract the sequence embeddings followed by a pooling and a linear layer. In an experiment, they showed that their Transformer-GAN achieves better performance compared to other Transformer models that were trained by maximizing the likelihood alone. 
Finally, Calliope was presented by \citet{valenti2021calliope}, and is a polyphonic music generation system based on adversarial autoencoders (AAE). 
A Transformer architecture is used for the encoder and the decoder while a multi-layer perceptron is used for the discriminator. 

Transformers have also been used for \textit{conditional} music generation, which is in essence what we propose in this paper. Except that instead of conditioning on key or emotion as is typically done is existing work, we condition on videos. 
\citet{makris2021generating} proposed an affective and controllable music generation system that is based on sequence-to-sequence architecture with long-short term memory (LSTM) and Transformer models. First, a sequence of musical attributes is given as conditions in the encoder stage. Then, this encoded feature is translated into lead sheet music (chords and melody) in the decoder stage. In experiments, they show that the Transformer has the best performance and can generate lead sheets that match desired valence levels. That same year, \citet{choi2021chord} presented a melody Transformer that is conditioned by chords and that can generate K-POP melodies. Their proposed model consists of two decoders: a rhythm decoder (RD) and a pitch decoder (PD). 
Another conditional music generation system was developed by~\citet{dai2021controllable}. In this system, the authors aim to model long-term structure through a hierarchical approach and a music representation called `Music Frameworks'. In this system, a full-length melody is created using a multi-step generative process with a Transformer-based model. 
The main idea is to adopt an abstract representation of basic rhythm forms, phrase-level basic melodies, and long-term repetitive structures. Then the melody is generated, conditioned on the basic melody, rhythm and chords in an auto-regressive manner. Their proposed architecture contains two elements: 1) an encoder which learns a feature representation of the inputs using two layers of Transformers and 2) a decoder which combines the last predicted note and the encoded representation as input and feeds them to one unidirectional LSTM to produce the final output which is the predicted next note. They demonstrated from a listening test that generated music pieces from their proposed model are rated as good as or better than the music pieces from human composers. 

In very recent work, Transformer architectures have also been used in Diffusion networks for monophonic symbolic music generation~\citep{mittal2021symbolic}, which further shows their ability to model music. 

In this work, we will be conditioning our Transformer network on video features. While many Transformer-based music generation models primarily focus on generating MIDI files, our proposed model generates chord sequences that match the video content. Our decision to work with chords is largely due to the lack of symbolic music-video datasets. Hence we first transcribed the chords from the audio. Since it is much more accurate to transcribe chords compared to polyphonic MIDI \citep{cheuk2023diffroll}, we opted to train on chords as this would propagate the least amount of error. In a post-processing step, these chords are appropriately arpeggiated and rendered expressively to further match the mood of the video and create a richer, more intricate sound. 

\subsection{Music Generation from Videos}
\label{sec:2.2}
The number of papers purely on music generation for video can be counted on one hand. However, there has been related work leading up to this. Below we expand on several papers related to generating music that serves as a narrative, be it for games, video, reconstructing instrument sounds, or other purposes. 

In the broader realm of narrative music, the concept of using musical cues to convey storytelling elements is essential~\citep{herremans2017functional}. Notably, the blending of music with other media, such as games and videos, has garnered substantial interest. For instance, game music, is often produced by cross-fading between audio files during transitions in game state~\citep{collins2008game}. One early attempt has been made by~\citet{johnson2006long} to dynamically generate music based on player interactions. \citet{casella2001magenta} proposed an abstract framework named MAgentA, distinct from Google's music generation project Magenta, which aims to enhance the generation of background music for video games. This framework focuses on producing `film-like' music that resonates with the emotional atmosphere of the in-game environment. The system achieves this by employing a cognitive model that captures the mood of the scene and translates it into musical elements. Any of the more recent music emotion conditioned music generation models~\citep{makris2021generating, guo2020variational, herremans2017morpheus} could be used in the future to match music with a game state based on mood in this way. However, to day, the limited existing research on music generation models for games uses mostly Markov models~\citep{prechtl2016adaptive, engels2015automatic} or procedural rules~\citep{plans2012experience}.

In films, the music component plays an important role in enhancing the emotional impact of visual narratives~\citep{parke2007quantitative}. Leveraging this, \citet{nakamura1994automatic} introduced a prototype system for generating background music and sound effects for short animated films. This system employs established rules from music theory to create harmonious elements such as harmony, melody, and rhythm for each scene. It takes into account variables like the mood's intensity and the musical key of the preceding scene to craft music that complements the visuals. In addition, the system employs an approach where sound effects are determined based on the distinct characteristics and the intensity of movements depicted on screen.

In the early 2000s, \citet{dannenberg2003sound} introduced a novel approach to music generation based on real-time video images. They explore the connection between visual imagery and sound by using video of light reflected from water to modulate sound spectra in real time. The authors address challenges in mapping video to sound and handling variations in light levels, showcasing the potential for video-based control over audio synthesis. 

More recently, \citet{di2021video} proposed CMT, a Controllable Music Transformer designed to generate background music for videos. This is one of the first deep learning models to try to generate music from video. In CMT, however, the video-music relationship, is solely rule-based and based on three key features. In contrast, our proposed method addresses these limitations by defining semantic, motion, scene offset, and emotion features extracted from the video, ensuring the generation of background music that aligns with the content of general videos. By considering these comprehensive features, our approach establishes a stronger connection between music and video. 

Finally, \citet{su2023v2meow} proposed V2Meow, a visually conditioned music generation system capable of producing high-fidelity music audio from silent videos. V2Meow utilizes pretrained visual features extracted from silent video clips to generate music audio waveforms. In addition, it provides control over the music style through supporting text prompts alongside video frame conditioning. V2Meow uses audio waveforms as the training input and output instead of symbolic music data (i.e., MIDI), which makes musical properties less explicit. It is thus harder for the system to understand musical relationships. As a result, the generated music lacks coherent musical ideas, logical progression, and nuanced musical expression. In contrast, our method uses symbolic music data (e.g., chords, key) so that it can more easily interpret musical structures, such as harmonic chord progressions and rhythmic patterns. By leveraging these rules, the generated music is more likely to exhibit a coherent and consistent musical structure, thus enhancing its quality.

A slightly different task, relates to reconstructing music from instrument performances that lack the accompanying audio~\citep{gan2020foley, koepke2020sight, su2020audeo, su2020multi}, such as generating piano music from a video of finger movements on the piano. Recently, related work has appeared that focuses on generating music for videos that feature dancing or human activities, emphasizing rhythmic relationships~\citep{su2021does, zhu2022quantized, zhu2022discrete}. These methods, however, necessitate additional motion annotations as input, limiting their applicability to specific videos and hindering their effectiveness for general videos encompassing diverse content.

In this work, we focus on generating new (symbolic) music, given a video as input condition. To achieve this, we first created a new dataset of music videos. In the next section, we discuss the details of this process.

\section{Dataset Creation}
\label{sec:3}
One of the key reasons that there are not many generative music for video systems out there, is the lack of symbolic music with video datasets. Given that the accuracy of music transcription systems is constantly growing \citep{cheuk2020impact, cheuk2021reconvat, cheuk2023diffroll}, especially chord transcription \citep{park2019bi}, we set out to design a novel way to create a dataset. This resulted in a new dataset, called MuVi-Sync, comprising both music features and video features extracted from a total of 748 music videos. Below, we describe the music and video features that we extract from this dataset. 

\subsection{Music Features}
\label{sec:3.1}
From the audio track of the music video, we extract four essential features: note density, loudness, chords, and key. These features play a crucial role in capturing the musical characteristics and composition of the audio. 
The chord and key features are utilized to train the decoder component of our Transformer model. This will help enable the generation of coherent and harmonically aligned chord sequences that match the video content. We leverage the note density and loudness features to train a post-processing model that transforms the chords into an arpeggiated MIDI file that is expressively rendered to better match the video. 

In the next subsection, we provide a detailed description how we obtained the note density, loudness, chord, and key features after extracting the audio tracks from the music videos. 

\subsubsection{Note Density}
\label{sec:3.1.1}
To compute the note density, we first converted the audio files from the music video to MIDI files. We used the OMNIZART Music Transcription library~\citep{Wu2021} to extract polyphonic MIDI files. We are aware that the transcription accuracy of this system is estimated to be around 72.50\% for frame-level F1-score and 79.57\% for note-level F1-score (pitch and onset) on the Configuration-II test set of the MAPS dataset~\citep{kelz2016potential}. However, for our purposes (post-processing), we believe that this is acceptable and will not influence the generated music too much. 

\begin{figure}[ht!] 
        \centering 
        \includegraphics[width=9 cm]{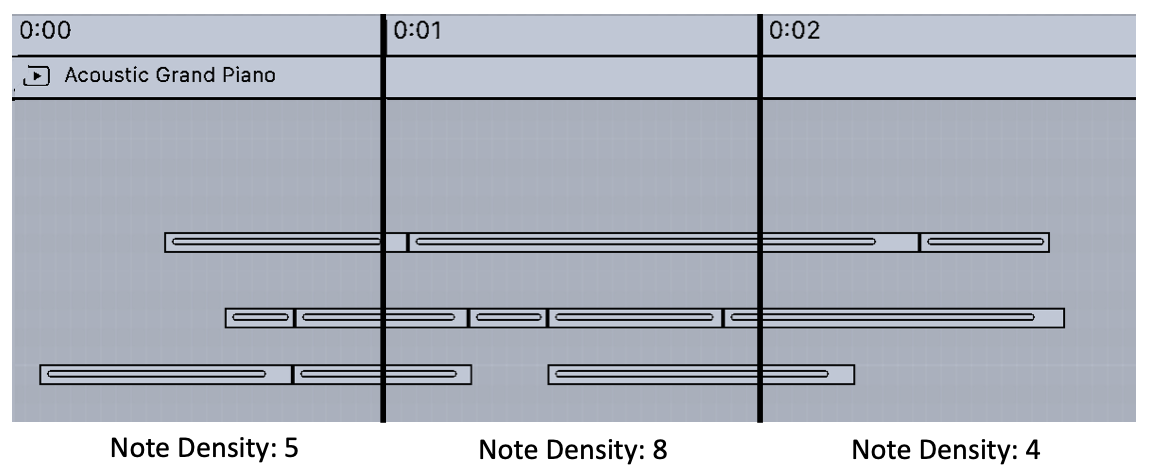} 
        \caption{Example of how note density is calculated for each 1s time window. In this example, the estimated note density values are 5, 8, and 4 for the intervals 0s to 1s, 1s to 2s, and 2s to 3s, respectively.}
        \label{fig:density} 
\end{figure}

For each transcribed MIDI file, we calculated the total number of notes in each 1 second interval, and used that value as the note density as shown in Figure~\ref{fig:density}.

\subsubsection{Loudness}
\label{sec:3.1.2}
To accurately estimate the loudness per second from the audio files, we used the audioop module from the Python standard library to calculate the root mean square (RMS) loudness. Subsequently, we convert the RMS loudness values to decibels (dB) and transform them to a 0-1 scale using the formulas below:

\begin{equation}
\label{eq:loudness1}
loudness_{dB} = 20 \cdot \log_{10}\left(\frac{loudness_{RMS}}{32767}\right)
\end{equation}

\begin{equation}
\label{eq:loudness2}
loudness_{transformed} = 10^{ loudness_{dB} / 20 }
\end{equation}
where 0 dB is represented as 1 and negative dB values are mapped to values between 0 and 1. This transformation process ensures a precise and consistent measurement of loudness given that decibels align more closely with the human perception of loudness, and mapping these values to a 0-1 scale further enhances the interpretability of the loudness value.

\subsubsection{Chords}
\label{sec:3.1.3}
We extracted chord sequences from the audio files by using the Transformer-based chord recognition model by~\citet{park2019bi}. One chord was detected per window of length 1s. This model achieves weighted chord symbol recall (WCSR) scores of 83.5\%, 80.8\%, 75.9\%, 71.8\%, 65.5\%, 82.3\%, and 80.8\%, respectively for the mir\_eval metrics~\citep{raffel2014mir_eval} of the Root, Thirds, Triads, Sevenths, Tetrads, Maj-min, and MIREX categories. These scores are acceptable or our purposes, especially since errors are often minor, e.g. confusing A minor versus C major or C major with C major seventh. The resulting chord sequences contain 13 different types of chords, including major, diminished, suspended, minor seventh (m7), minor, suspended second (sus2), augmented, diminished seventh (dim7), major sixth (maj6), half diminished seventh (hdim7), seventh (7), minor sixth (m6), and major seventh (M7).

\begin{figure}[ht!] 
        \centering 
        \includegraphics[width=12 cm]{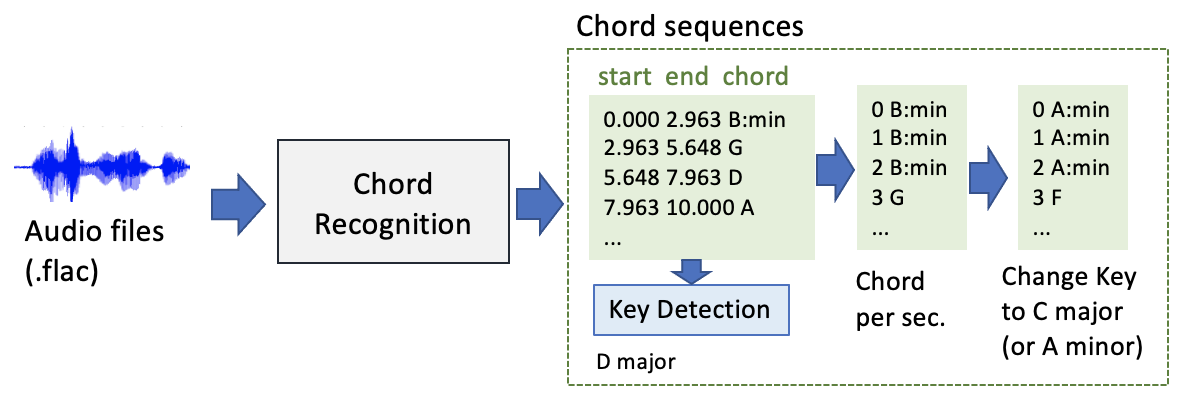} 
        \caption{Chord recognition and normalization procedure. Chord sequences, along with their respective start and end times, are identified from the audio file using a chord recognition model. Subsequently, the detected chords are reformatted to a one-chord-per-second representation. Depending on whether the recognized key is major or minor, the song's chords are transposed to either C major or A minor.}
        \label{fig:chord} 
\end{figure}  

The chord sequence for each file was normalized as per the detected key (see next subsection). Depending on whether the detected key was major or minor, we transposed the song's chords to C major or A minor, respectively. Figure~\ref{fig:chord} shows the overall procedure to extract and normalize the chords from the music video.

\begin{figure}[ht!] 
        \centering 
        \includegraphics[width=12 cm]{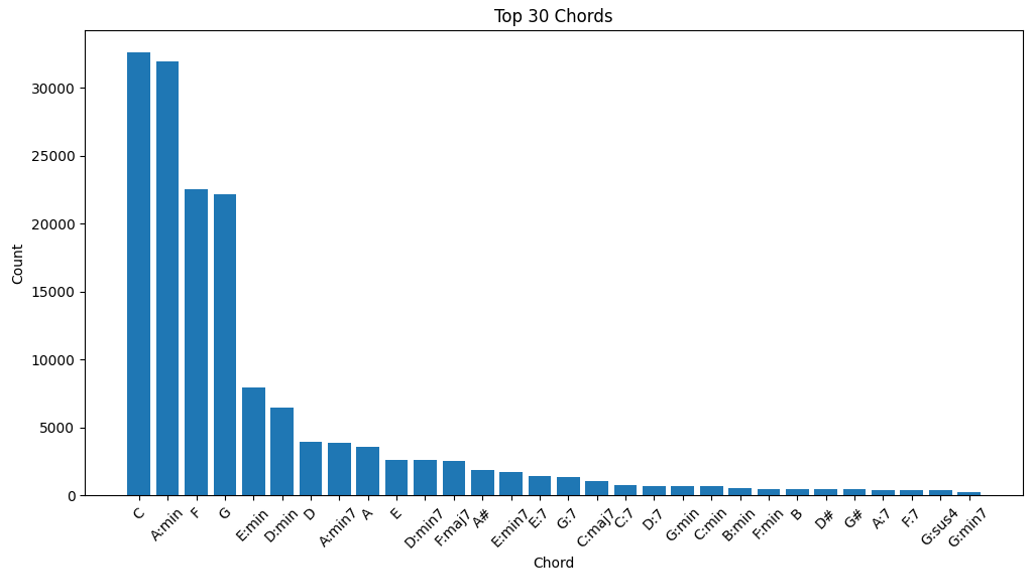} 
        \caption{Top 30 normalized chords (to either the C major or A minor key) in our dataset.} 
        \label{fig:top30chord} 
\end{figure}

Figure~\ref{fig:top30chord} shows the top 30 normalized chords (to either the C major or A minor key) in our dataset. Unsurprisingly, the most popular chords are the major or minor root chord, followed by the IV, and V. 

\subsubsection{Key}
\label{sec:3.1.4}
After extracting the chord sequences, we proceeded to convert them into MIDI files using simple music theory. For instance, a C major chord translates to the notes C, E, and G, whereas a C minor 7th chord corresponds to the notes C, Eb, G, and Bb. We use the start and end times of each chord to precisely map the duration of each chord to notes in the MIDI file. We can then use the MIDI files to determine the key of each song. Leveraging the music21 library's key detection functionality~\citep{cuthbert2010music21}, we employed three commonly used key finding algorithms: Krumhansl-Schmuckler~\citep{krumhansl2001cognitive}, Temperley-Kostka-Payne~\citep{temperley2007music}, and Bellman-Budge~\citep{bellmann2006determination}. 

\begin{figure}[ht!] 
        \centering 
        \includegraphics[width=12 cm]{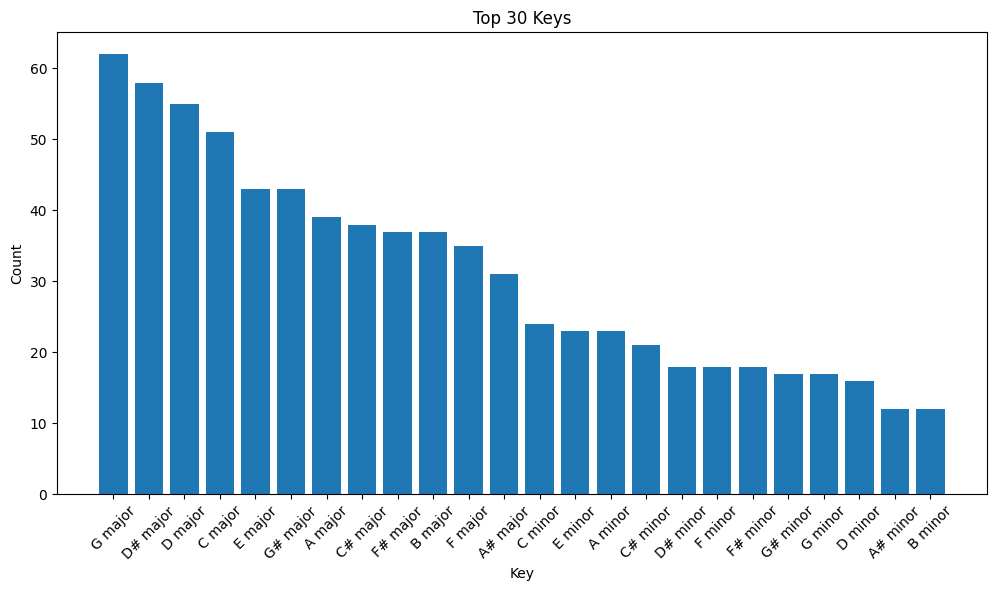} 
        \caption{Top 30 keys in our dataset (before chord normalization).} 
        \label{fig:top30key} 
\end{figure}

Each algorithm provides a candidate key for a given song based on different musicological principles. The choice of these specific algorithms lies in their distinct methodologies and proven effectiveness in diverse musical contexts \citep{kania2022comparison}. Krumhansl-Schmuckler relies on cognitive principles related to tonal hierarchies, Temperley-Kostka-Payne integrates probabilistic models, and Bellman-Budge incorporates harmonic stability considerations. By employing this combination, we aim to capture a broad spectrum of musical features that contribute to key identification. To consolidate the results obtained from the three algorithms, we implemented a voting method. This approach involves each algorithm `casting a vote for the predicted key, and the final predicted key is determined based on the most commonly selected key among the algorithms. This ensemble decision-making process enhances the reliability and accuracy of the predicted key for each song.
Figure~\ref{fig:top30key} shows the top 30 keys detected in our dataset.

\subsection{Video Features}
\label{sec:3.2}
If we were to use the raw video frames directly as conditional input to our generative music model, it would be challenging for the model to effectively learn the correspondence between disparate modalities. Therefore, we extract meaningful features from video as intermediate representations to simplify the learning process. We extract semantic features, emotion, scene offset, and motion features to guide the music generation model. We used one video frame for each second of video to extract the below features. The extraction process for each of these features is described below. 

\subsubsection{Semantic features}
\label{sec:3.2.1}
We harnessed the capabilities of CLIP (Contrastive Language-Image Pretraining)~\citep{radford2021learning}, a powerful pretrained model, as a feature extractor. This model enabled us to encode the raw video frames into semantic feature tokens without the need for fine-tuning. We utilized CLIP to extract latent features from each video frame. These extracted latent features encompass a wide range of video semantics, including serene beach scenes, adventurous outdoor activities, and bustling city streets. 

\subsubsection{Emotion}
\label{sec:3.2.2}
To estimate the emotions expressed in a (muted) video on a per-second basis, we employ the CLIP model~\citep{radford2021learning}. This model has been trained on an extensive dataset containing 400 million image-text pairs, providing it with a robust understanding of the relationship between images and text.

\begin{figure}[h!] 
        \centering 
        \includegraphics[width=12cm]{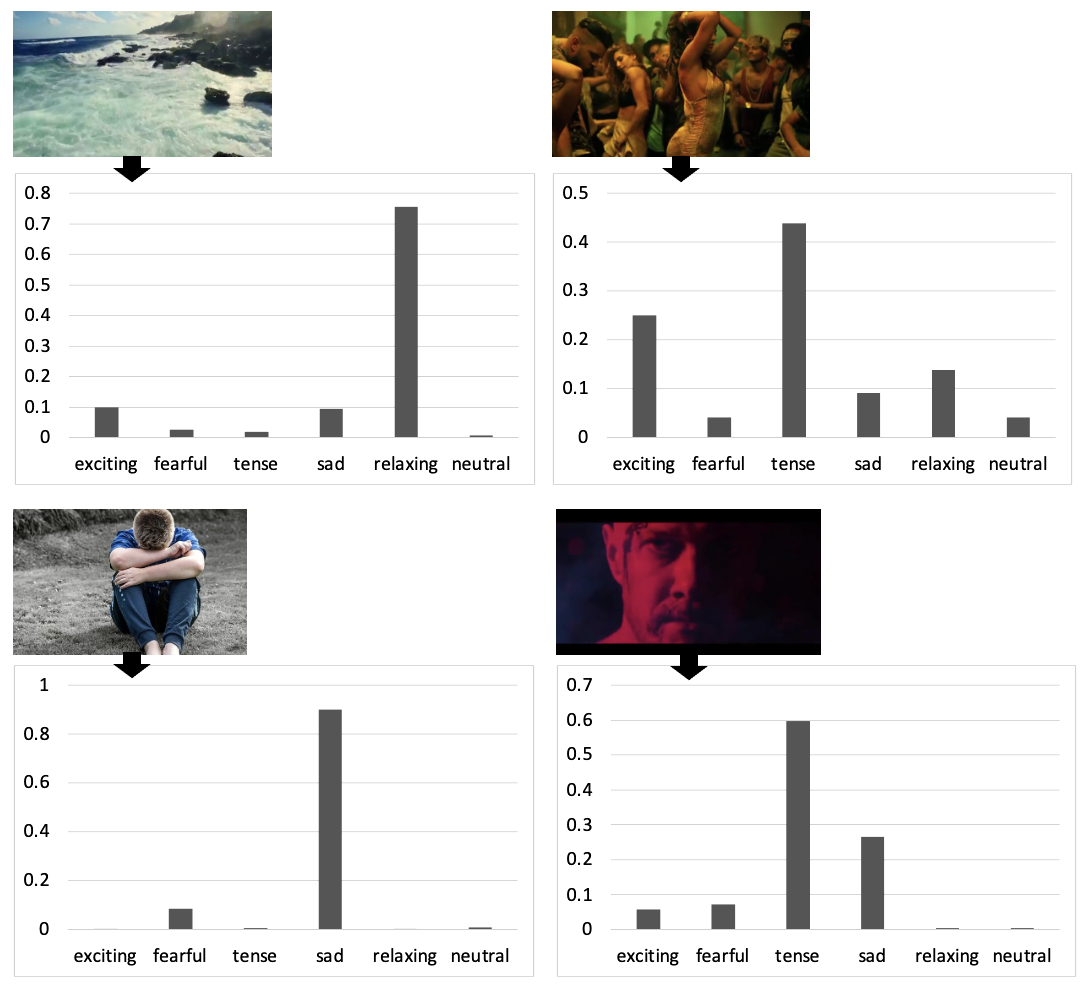} 
        \caption{Examples of the CLIP probability results for each of the six emotion classes.} 
        \label{fig:emotion} 
\end{figure}

In our case, CLIP serves a dual purpose. Firstly, it extracts semantic features from the video, as discussed in the previous subsection. Secondly, we include its probabilities for six emotion classes. By leveraging its pre-trained knowledge acquired through exposure to diverse image-text pairs during training, CLIP can provide the probability distribution of different emotion classes (`exciting,' `fearful,' `tense,' `sad,' `relaxing,' and 'neutral') for each frame in the video as shown in Figure~\ref{fig:emotion}. The selection of these emotion classes was based on the MVED dataset~\citep{pandeya2021deep} which includes 5,743 music video segments annotated with six emotion labels (`exciting,' `fearful,' `tense,' `sad,' `relaxing,' and 'neutral').

To obtain these values for each 1s video, we employ a smoothing window with a size of 5 for each of the six emotion probability time series. A smoothing window is a computational tool used to minimize short-term fluctuations or noise in data. This window moves across the data, and at each position, it computes the average of the values within the window. The result is a smoothed version of the original data, where abrupt changes or minor fluctuations are mitigated, providing a clearer representation of the underlying trends in the emotion probabilities over time. Since we extract one frame per second, this means that to calculate the emotion, we look back 5 seconds in time. Essentially, this method helps reveal the broader patterns by averaging out the smaller, potentially noisy variations.

\subsubsection{Scene offset}
\label{sec:3.2.3}
\begin{figure}[ht!] 
        \centering 
        \includegraphics[width=12 cm]{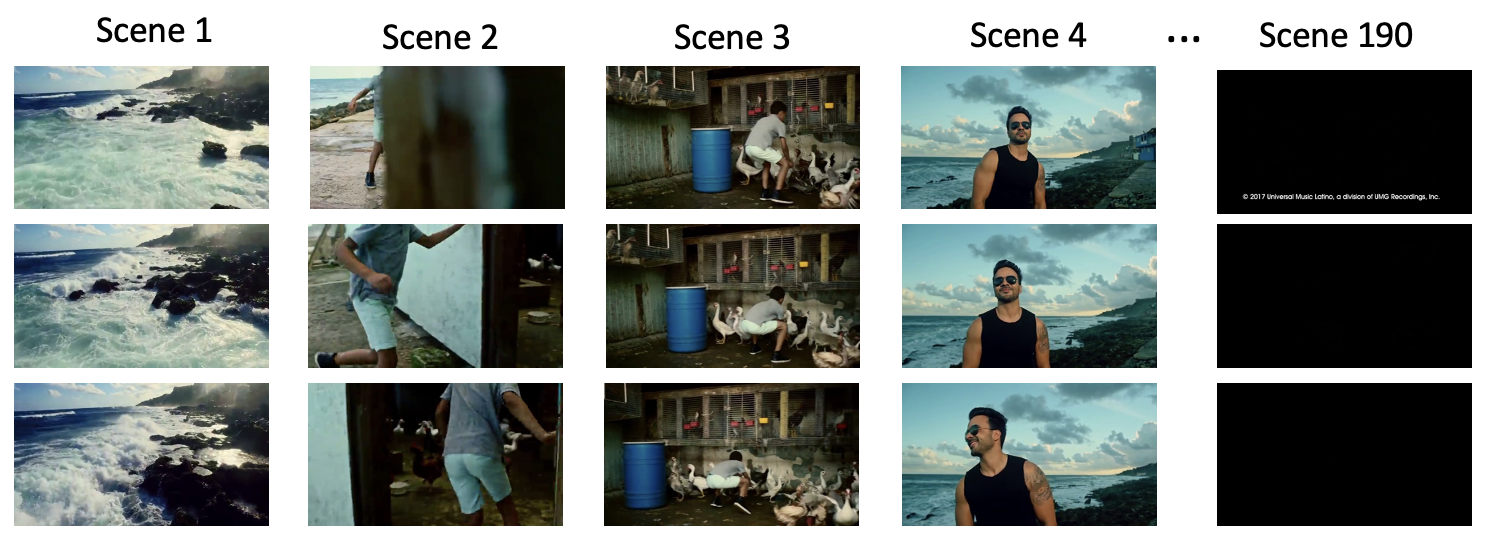} 
        \caption{Examples of video frames that belong to the same detected scene. } 
        \label{fig:scene} 
\end{figure}

We used the PySceneDetect library~\citep{castellano2018pyscenedetect} to accurately detect shot changes in videos. Figure~\ref{fig:scene} shows a few examples of the scene detection results. Instead of directly utilizing scene IDs as a feature to incorporate scene change information, we chose to calculate scene offsets based on the detected scene IDs. By introducing a scene offset value that initiates at 0 and progressively increments until the next scene change, we effectively capture the relative position of each frame within a scene. This approach allows us to implicitly encode the scene change information by considering the temporal distance between frames within and across scenes.

\subsubsection{Motion}
\label{sec:3.2.4}
\begin{figure}[h!] 
        \centering 
        \includegraphics[width=12 cm]{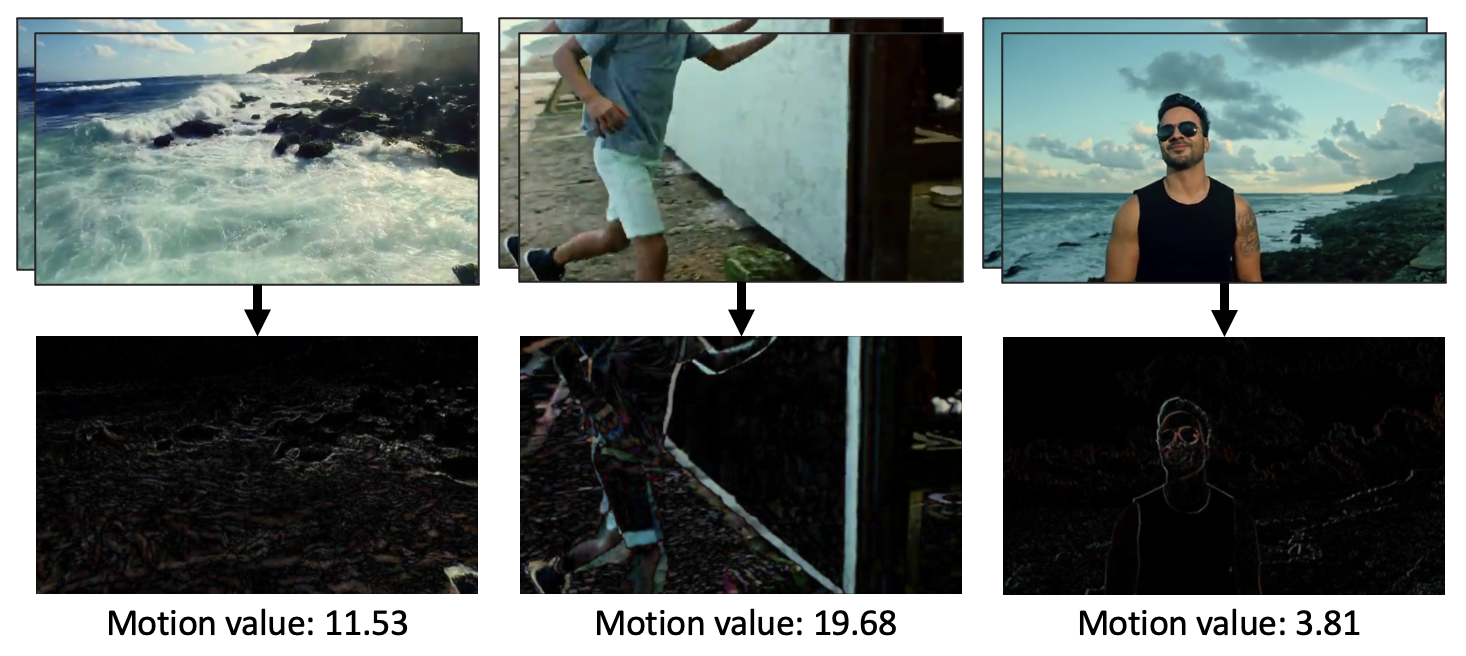} 
        \caption{Examples of RGB difference and the resulting motion value. For static scenes such as the portrait shot on the right, the motion value is low. } 
        \label{fig:motion} 
\end{figure}

To estimate motion or changes in the visual content of the video, we first computed the RGB Difference between the current frame and the preceding frame within each one-second interval. This process involves calculating the absolute difference in color values for corresponding pixels across the Red, Green, and Blue channels independently. Following this, we determined the mean of all pixel values in the resulting RGB difference image. This computed average value serves as our motion value feature, effectively representing the overall disparity between corresponding pixels in the two frames for the specified one-second interval. Figure~\ref{fig:motion} shows examples of RGB Differences and the resulting motion values

\section{Proposed Video2Music framework}
\label{sec:4}
\begin{figure}[h!] 
        \centering 
        \includegraphics[width=14 cm]{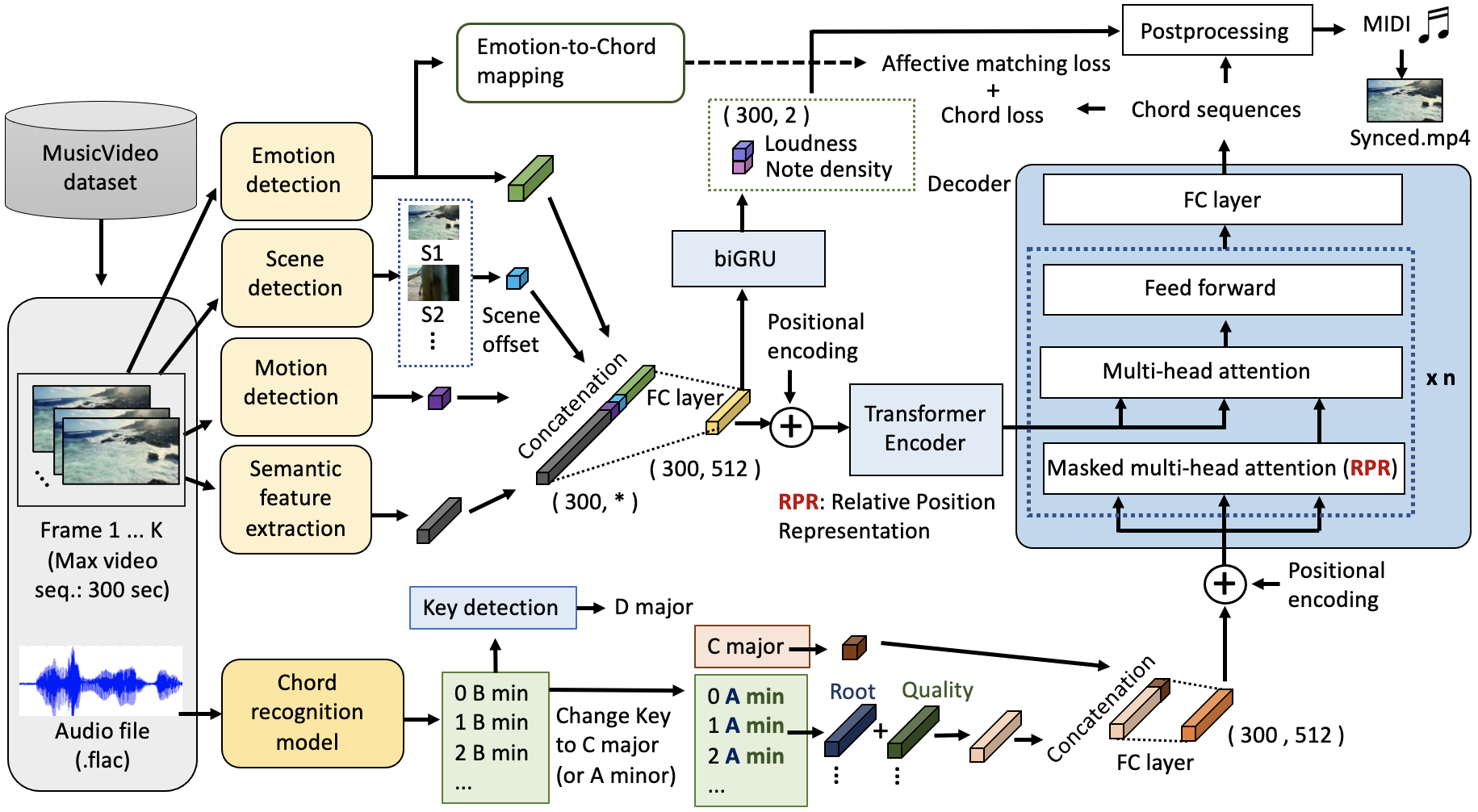} 
        \caption{Our proposed Video2Music framework for generating music based on video. Video features (scene, motion, emotion, and semantic) are extracted from music videos as well as audio features (chords and key). These features are then concatenated and processed through a Transformer-based model to generate expressive chord sequences. A post-processing step estimates note density and loudness for a dynamic MIDI output using biGRU. This allows us to enhance the musical expressiveness of the final audio to better match the input video.} 
        \label{fig:framework} 
\end{figure}

The overall framework of the proposed music generation system is shown in Figure~\ref{fig:framework}. First, we extract both video features (scene, motion, emotional flow, and semantic) and music features (chord and key) from music videos for every second, as described in the previous section. We then concatenate these video features into a 2-dimensional sequence, and apply a fully-connected layer to create the final video embedding vector. The latter is further fed into the Transformer encoder. The key serves as a conditional input feature, and is concatenated with the chord embedding. This is crucial because, a song's key greatly influences its chord progressions. The resulting chord embedding is calculated by summing the chord root (e.g., `A') embedding and chord type (e.g., `minor') embedding. This fusion forms a comprehensive music embedding vector, which is fed to the Transformer decoder to generate the chord sequences that match the input video.

In order for the model to attend to the order of the sequence of music and video features, we add positional encodings to both the music and video embedding vectors, before feeding it to the encoder and decoder, respectively. We use the relative position representation (RPR) introduced in Music Transformer~\citep{huang2018music} for the masked multi-head attention module of our Transformer decoder. Our decoder learns to predict the next chord sequences given input video features as well as the previous chords.

Finally, a post-processing step uses a regression model based on bi-directional Gated Recurrent Units (biGRU) to estimate the note density and loudness based on the video features. This way, the resulting MIDI file dynamically adjusts the rendering of the generated chords, introducing variations in rhythm and volume for a more expressive musical output that better matches the video.

\subsection{Affective Multimodal Transformer}
\label{sec:4.1}
The Transformer model~\citep{vaswani2017attention} is an encoder-decoder based auto-regressive generative model, which was originally designed for machine translation applications. We adopt the basic architecture of this model and consider our task as a video to chord translation problem. The core architecture of our proposed Affective Multimodal Transformer (AMT) comprises of two key components: a Transformer encoder responsible for capturing video features extracted from the video, and a Transformer decoder that generates chord sequences by intelligently leveraging the context of preceding chords as well as employing a cross-attention mechanism that fuses information from both the musical and visual modalities.

\subsubsection{Input Representations}
\label{sec:4.1.1}
An effective input representation is essential for seamlessly integrating musical and visual information into the Transformer model. For audio, after extracting the chords at every second of the audio tracks, we disassemble them into two essential components: the chord root (e.g., C, D) and the chord type (e.g., minor, major, diminished). Each component is encoded as a one-hot vector. Then, we apply an embedding function to both of these vectors. These embeddings are then summed, producing a comprehensive chord embedding vector that encapsulates both the chord root and chord type information.

We concatenate this chord embedding vector with a 1-dimensional vector that represents the key of the song. Given the key normalization (see Section~
\ref{sec:3.1.3}), this vector can simply contain the value 0 for minor and 1 for major. Finally, this concatenated vector is passed through an embedding layer, yielding a final input embedding vector with a dimensionality of 512. This enriched embedding vector becomes part of the input for training our Transformer decoder, together with the video embedding vector, to generate sequences that match with the video's content. This entire process is represented by the equation:

\begin{equation}
\text{Input}_{\text{music}}^{t} = PE(E_{\text{chord}}(\text{concat}(k, E_q(C_q^t) + E_r(C_r^t))))
\end{equation}
where \( \text{Input}_{\text{music}}^{t} \) represents the input music vector at a given time $t$, \( \text{concat}() \) represents the concatenation function, and \( PE() \) represents the positional encoding function which is used to inject information about the position or order of elements in a sequence into the representation of those elements. The \( k \) represents the key vector that has a value of either 0 (minor) or 1 (major), \( C_q^t \), \( C_r^t \) represents the one-hot chord type vector and one-hot chord root vector at a given time $t$, respectively, and \( E_q() \), \( E_r() \) ,\( E_{chord}() \) represent the embedding functions for chord type, chord root, and chord respectively. 
Embedding functions are a way to represent categorical variables as continuous vectors in a high-dimensional space.

We follow a similar procedure for representing the video input, represented by the equation:

\begin{equation}
\label{eq:input_video}
\text{Input}_{\text{video}}^{t} = PE(FC(\text{concat}(V_{\text{scene}}^t, V_{\text{motion}}^t, V_{\text{emo}}^t, V_{\text{sem}}^t)))
\end{equation}
where \( \text{Input}_{\text{video}}^{t} \) represents the input video vector at a given time $t$, and \( V_{\text{scene}}^t \), \( V_{\text{motion}}^t \),\( V_{\text{emo}}^t \), and \( V_{\text{sem}}^t \) represent the scene offset vector, motion vector, emotion vector, and semantic vector respectively at a given time $t$. Finally, \( FC() \) represents a fully connected layer that maps the concatenated video feature vector into a 512-dimensional space.

\subsubsection{Transformer Encoder}
\label{sec:4.1.2}
In the process of encoding input vectors that represent videos, denoted as $\text{Input}_{\text{video}}$, a Transformer Encoder with $L$ layers is utilized. Each layer $l$ within the range $1 \leq l \leq L$ takes the current contextual representation $H^{(l-1)}$ and transforms it into the subsequent output $H^{(l)}$ through the Transformer mechanism. 
Each layer consist of two primary sub-layers: Multi-Head Self-Attention and Position-wise Feed-Forward Networks. Within the Multi-Head Self-Attention sub-layer, the core mechanism is the Self-Attention Mechanism~\citep{vaswani2017attention}, which computes the attention scores as follows:

\[
\text{Attn}(Q, K, V) = \text{softmax}\left(\frac{QK^T}{\sqrt{D_k}}\right)V
\]

where $Q$, $K$, and $V$ are linearly transformed versions of the input, and $\sqrt{D_k}$ is the scaling factor. The result is a weighted sum of values $V$ based on the compatibility of queries $Q$ and keys $K$. 

The initial contextual representation is $H^{0}$, set to be the input vector $\text{Input}_{\text{video}}$. After the final layer ($L$), we obtain the final contextualized representations, denoted as $H^{(L)}$ from $\text{Input}_{\text{video}}$, which are subsequently passed into the multimodal cross-attention module within the Transformer Decoder for further processing.

\subsubsection{Transformer Decoder}
\label{sec:4.1.3}
The Transformer Decoder architecture~\citep{vaswani2017attention} is employed in our Affective Multimodal Transformer (AMT) to generate chord sequences with long-term dependencies. In the Decoder, music features are processed using a cross-attention module following a masked multi-head self-attention module. Simultaneously, the encoded video features from the Transformer Encoder are used as keys and values.

Specifically, let $\text{Input}_{\text{video}} \in \mathbb{R}^{C \times H}$, where $C$ represents the total number of Chord events contained in a video clip, and $H$ is the hidden dimension. The Transformer Decoder's goal is to predict a sequence of Chord events $\text{Output}_{\text{chord}} \in \mathbb{R}^{C \times L}$ where $L$ is the vocabulary size of chord events. At each time step, the Decoder takes as input the previously generated feature encoding over the chord event vocabulary and the visual features, to predict the next chord event.

In contrast to the standard Transformer model, which uses positional sinusoids for timing information, we incorporate relative position representations~\citep{shaw2018self}. These representations explicitly encode the distance between tokens in a sequence, a crucial consideration for music applications~\citep{huang2018music}. We adopt a strategy similar to~\citet{huang2018music} to jointly learn ordered relative position embeddings $R$ for all possible pairwise distances among pairs of query and key positions within each attention head:

\[
\text{Attn}_{\text{relative}}(Q, K, V) = \text{softmax}\left(\frac{QK^T + R}{\sqrt{D_k}}\right)V
\]

For our Transformer Decoder, we first use a masked self-attention module that incorporates relative position embeddings to encode input chord events. In this module, queries, keys, and values are all derived from the same feature encoding and the attention mechanism only takes into account the current and preceding positions, preserving the sequential nature of the data. The output of the masked self-attention module, along with the output of the Transformer Encoder processing the input video features, is then passed into a multi-head attention module, computed as follows:

\[
\text{Attn}_{\text{cross}}(Q_{\text{dec}}, K_{\text{enc}}, V_{\text{enc}}) = \text{softmax}\left(\frac{Q_{\text{dec}}K_{\text{enc}}^T}{\sqrt{D_k}}\right)V_{\text{enc}}
\]
where $Q_{\text{dec}}$ represents the query matrix derived from the Decoder's hidden states, $K_{\text{enc}}$ represents the key matrix derived from the Encoder's hidden states (used for cross-attention), and $V_{\text{enc}}$ represents the value matrix derived from the Encoder's hidden states. This cross-attention mechanism enables the Decoder to focus on relevant information from the input video features while generating the next chord event, thus facilitating the modeling of music events and dependencies.

Following the cross multi-head attention layer, the Transformer Decoder incorporates a pointwise feed-forward layer. This layer plays a pivotal role in further transforming the encoded information. Subsequently, the output from the feed-forward layer is passed through a linear transformation followed by a softmax activation function. This step is essential for generating probability distributions over the vocabulary. At this step, each token in the vocabulary receives a probability score, indicating the likelihood of it being the next token in the output chord sequence. We include a few hyperparameters to further improve this selection. For instance, we set the maximum number of repeated chords to 2 and the maximum number of repeated silences also to 2. In case these constraints are met, the chord with the second highest probability is selected. 

\subsubsection{Affective Matching Loss Function}

\label{sec:4.1.4}
In our proposed Affective Multimodel Transformer (AMT) architecture, the total loss is calculated as the weighted sum of the loss related to chords, \( L_{\text{chord}} \), and the loss related to the emotion of the resulting chords, \( L_{\text{emo}} \), as follows:

\begin{equation}
L_{\text{total}} = \lambda L_{\text{chord}} + (1- \lambda) L_{\text{emo}}
\end{equation}
where \( \lambda \) represents a weighting factor that determines the relative importance of the two individual loss components in the total loss. First, the \( L_{\text{chord}} \) can be calculated as the cross-entropy between the soft targets of the model estimated by the softmax function, and the ground-truth labels as follows:

\begin{equation}L_{\text{chord}}(y^{\text{chord}},z) = -\sum_{i=0}^{M}y_i^{\text{chord}}\text{log}\left (\frac{ exp(z_{i})}{\sum_jexp(z_{j})}\right )\label{eq1}\end{equation}
where $M$ is the total number of classes, \(  y ^{\text{chord}} \) is a one-hot vector which represents the ground-truth label of the training dataset as 1, and \(  z_{i}  \) is the logit (the output of the last layer) for the i-\textit{th} class of the model. 

Secondly, the \( L_{\text{emotion}} \) is defined as follows:

\begin{equation}
\label{eq:loss_emotion}
L_{emo}(y^{emo},z) = - \frac{1}{M}  \sum_{i=0}^{M}y_i^{emo}\text{log}( \sigma (z_i) ) +(1 - y_i^{emo})\text{log}( 1-\sigma (z_i) ) 
\end{equation}
where \( y^{emo}  \) is a ground-truth emotion vector that corresponds to the chord type attributes (e.g. minor, minor 7th, see Section~\ref{sec:3.2.2}) associated with a specific emotion of video frame predicted by CLIP model~\citep{radford2021learning}. The matching chord type positions of this vector, which has a similar format to the output vector of the decoder, that belong to the predicted emotion are activated. These chord type attributes are compared to the generated chord qualities. Then, \( y_i^{emo} \) is a $i$-th element of \( y_i^{\text{emo}} \), and \( \sigma (z_i) \) is the sigmoid function, which transforms the logit \(  z_{i}  \) into a value between 0 and 1. 

To obtain the chord attributes that match the emotion of the video, we use the following procedure: first, we use CLIP to obtain a probability for each of our five emotion categories for the video fragment. The choice of these five emotions (exciting, fearful, tense, sad, relaxing) for mapping to corresponding chords is grounded in the MVED dataset~\citep{pandeya2021deep} as discussed in Section~\ref{sec:3.2.2}. Then, we take the emotion with the highest probability, and use Table~\ref{tab:emochord} to find the matching chord attributes. If an emotion has multiple chord attributes, this vector can be multiple-hot. For instance, if the highest predicted emotion from the video is `sad', the elements in \( y^{emo} \) that correspond to the attributes `min7', `min' and `sus2' are set to 1. 

\begin{table}[ht!]
\scriptsize
\caption{Mapping of emotions with associated chord types based on the insights of professional musicians, music theory~\citep{chase2006music}, and music psychology~\citep{schuller2010determination}. }
    \centering
	\label{tab:emochord}
        \begin{tabular}{lccccccccccc}
        	\toprule
        	    Emotion & maj & dim & sus4 & min7 & min & sus2  & dim7 & maj6
                & hdim7 & 7 & maj7 \\
        	\midrule
                Exciting & \checkmark & & \checkmark & & & & & & & \checkmark &   \\
                Fear & & \checkmark & & \checkmark & & & \checkmark & & \checkmark & &   \\
                Tense & & \checkmark & \checkmark & \checkmark & & & & & & \checkmark &   \\
                Sad & & & & \checkmark & \checkmark & \checkmark & & & & &   \\
                Relaxing & \checkmark & & & & & & & \checkmark & & & \checkmark  \\
            \bottomrule	
        \end{tabular}
\end{table}

Table~\ref{tab:emochord} was derived from insights of professional musicians and music theory~\citep{chase2006music}, and augmented with work from music psychology~\citep{schuller2010determination, makris2021generating}. \citet{schuller2010determination} provides interesting insights on the connection of chord types with emotions. We base ourselves on their results to populate the table. For instance, in their results, a maj7 chord is related to `Romance, softness, jazziness, serenity, exhilaration, tranquillity', which we find close to our emotion category `relaxing', and a dim7 chord is labeled as `Fear, shock, spookiness, suspense', which clearly falls into our category `fear'. Other, less clear or missing, mappings where deliberated with professional musicians. And example is the sus4 chord, which \citet{schuller2010determination} labels as `Delightful tension'. In our table, this chord type maps to both the `tense' as well as `excited' emotion categories. 

\subsection{Post-processing to generate MIDI}
\label{sec:4.2}
After the model generates the chords, we perform post-processing to obtain a playable MIDI file. In this phase, we fine-tune the music's attributes, ensuring its harmonious fusion with the accompanying video. This section delves into the key steps we take to achieve this synchronization.

\subsubsection{Loudness and note density estimator}
\label{sec:4.2.1}
Leveraging the same input video embedding vector used for the Transformer encoder of our Affective Multimodal Transformer model, we trained regression models to jointly predict the extracted loudness and note density from the original audio. These predicted values are subsequently used to select chord arpeggiation patterns and perform MIDI velocity adjustments. This results in music that boasts nuanced rhythmic variations and dynamic intensities, synchronized with the video's mood and tempo.

We explored five difference regression models for estimating note density and loudness: 1) Fully-connected layer (FC) which was implemented with two linear layers: the first transforms the input video feature dimension to 512 with a ReLU activation function, and the second produces a single output unit for regression, 2) Long short-term memory (LSTM) which employed a dual-layer LSTM structure, with each layer consisting of 64 nodes, 3) Bi-directional LSTM (Bi-LSTM) which integrates bidirectionality, resulting in 128 nodes ($2\times64$) to capture information from both directions, 4) Gated recurrent units (GRU) which employed a dual-layer GRU structure, with each layer consisting of 64 nodes, and 5) Bi-directional GRU (Bi-GRU), which integrates bidirectionality, resulting in 128 nodes ($2\times64$). We use RMSE (Root Mean Square Error) as the metric to evaluate the performance of our regression models. This metric captures the average magnitude of the differences between the predicted and actual values in a regression problem: 

\begin{equation}
RMSE = \sqrt{ \sum\nolimits_{i=1}^{n} \frac {(\hat{y}_i - y_i)^{2}}{n} }
\end{equation}
where \( n \) represents the number of samples, \( \hat{y}_i \) is the predicted values of \textit{i}-th sample, \( y_i \) is the actual values of \textit{i}-th sample.

Table~\ref{tab:rmse} shows the Root Mean Square Error (RMSE) of note density and loudness for the different regression models. As can be seen from the table, the Bi-GRU model performs best. Hence, in our Video2Music framework, we adopt the Bi-GRU model to estimate both note density and loudness during post-processing.

\begin{table}[ht!]
\caption{Root Mean Square Error (RMSE) for estimating note density and loudness based on video features, for different regression models.}

    \centering
	\label{tab:rmse}

    \begin{tabular}{lcc}	
	\toprule
	    Model & RMSE (Note density) & RMSE (Loudness)  \\
	\midrule

FC & 4.7314 & 0.0877 \\

LSTM & 4.6247 & 0.0892  \\

Bi-LSTM & 4.5337 & 0.0882   \\

GRU & 4.6030 & 0.0888   \\

Bi-GRU & \textbf{4.5030} & \textbf{0.0876}   \\
            \bottomrule	
        \end{tabular}
\end{table}

\subsubsection{Chord arpeggiation based on note density}
\label{sec:4.2.2}
To make the resulting music rhythmically interesting, we perform arpeggiation to the generated chords. Arpeggiation spreads out the notes of a chord over time, in patterns that may repeat, often in a progressively upwards or downward order~\citep{kamien1988music}. We have selected five popular chord arpeggiation patterns as shown in Table~\ref{tab:arp} and select the best fitting one based on the note density estimation. Higher note densities will be assigned faster patterns like 5th Pattern in Table~\ref{tab:arp}  and vica versa. For example, if the note density is predicted to be high, we might play the chords with a lot of fast notes to make the music energetic (e.g. 5th Pattern in Table~\ref{tab:arp}). On the other hand, if the note density is predicted to be low, we might play the chords with slower and more spaced-out notes to create a softer and relaxed feeling (e.g. first Pattern in Table~\ref{tab:arp}). This step adds a rhythmic touch to the music, aligning it with the video's pacing and mood.

\begin{table}[ht!]
\caption{Arpeggiation patterns for different note density levels. For instance, for a C major chord which consist of the notes C4 (C note in the fourth octave), E4, G4, and C5, the `1', `2', `3', and `4' in the arpeggiation pattern refers to the note C4, E4, G4, and C5, and the symbol `*' represents a silent note, contributing to the rhythmic structure. The timestep of arpeggiation patterns for each chord ranges from 1/8 sec to 8/8 sec. In the scenario where the previous chord is the same as the current one (e.g., \dots, C, \textbf{C}), the next arpeggiation pattern is applied to the timesteps 9/8 sec to 16/8 sec.
}

    \centering
	\label{tab:arp}
    \begin{tabular}{lcc}	
	\toprule
	    Note density level & Arpeggiation patterns   \\
	\midrule
1 (very low, $\leq$ 5 notes)  & 1 * * * 2 * * * 
3 * * * 4 * * *  
  \\
2 (low, 6-10 notes) & 1 * 2 * 3 * * * 4 * 2 * 3 * * *
   \\
3 (moderate, 11-15 notes) & 1 * 2 * 3 * 4 * 3 * 2 * 3 * 4 *
    \\
4 (high, 16-20 notes) & 1 2 3 2 4 * 3 * 2 1 2 3 4 * 3 * 
    \\
5 (very high, $\geq$ 21 notes) & 1 2 3 2 4 3 2 3 2 1 2 3 4 3 2 3 
  \\
    \bottomrule	
    \end{tabular}
\end{table}

\subsubsection{Velocity estimation based on loudness}
\label{sec:4.2.3}
We recognize that the music's volume should synchronize with the emotional intensity and visual dynamics of the video. To achieve this synchronization, we convert the predicted loudness based on the video features (with the model described above), into a parameter known as MIDI velocity, which governs the perceived loudness of the notes in the music. 
The conversion is achieved through a linear mapping procedure, where the predicted loudness values, ranging from 0 to 1, are translated into corresponding MIDI velocity values within the defined MIDI velocity range of 49 to 112.

By establishing a connection between loudness levels and visual features, we hope to forge a cohesive link between the auditory and visual elements. As the video becomes more intense, the music can respond by growing louder, and as the video becomes more tranquil, the music may adopt a softer demeanor.

\subsection{Web User Interface}
\label{sec:4.3}
\begin{figure}[h!] 
        \centering 
        \includegraphics[width=14 cm]{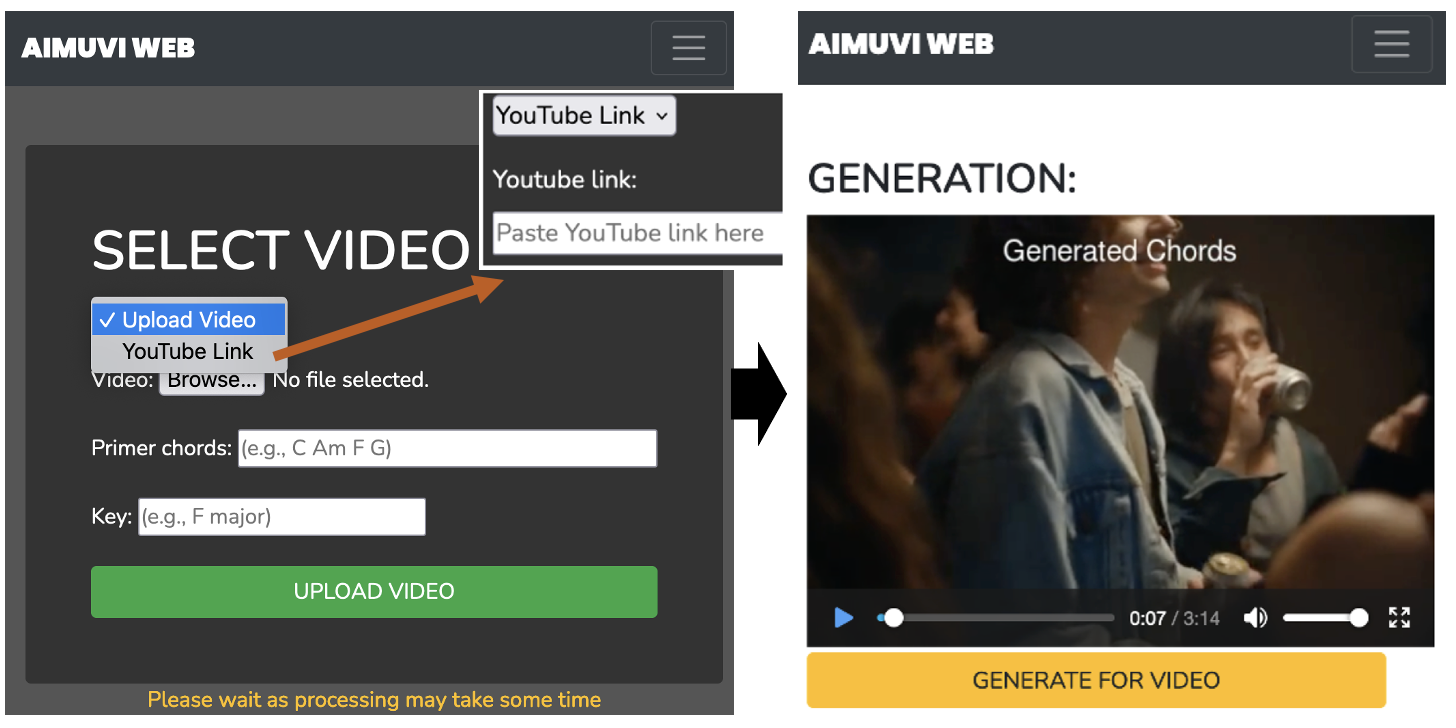} 
        \caption{Interactive web user interface. Left: Input page where users can specify the video, as well as the primer chord, and key. Right: Output page displaying the new video with generated music.}
        \label{fig:web} 
\end{figure}

To make our music generation accessible and user-friendly, we have incorporated our models into an intuitive web interface using the Django web framework~\citep{django} as shown in Figure~\ref{fig:web}. This interface empowers users with the following capabilities:

\begin{enumerate}
    \item \textbf{Video Selection}: Users can effortlessly select their desired video by either uploading it directly or providing a YouTube link.
    \item \textbf{Key and Chord Progression Specification}: Users can specify the key and optional seed chord progression (e.g., C Am F G) as a primer.
\end{enumerate}

Once the video is chosen and uploaded, our model processes the video and generates a matching audio file. To achieve this, the MIDI file is rendered through a FluidR3 General MIDI soundfont. Finally, the original video stream is synced with the newly generated audio using MoviePy~\citep{burrows2021moviepy} . The resulting music video .mp4 file is offered as a download. 
In addition, a live demo of the model is available on Huggingface Spaces\footnote{\url{https://huggingface.co/spaces/amaai-lab/video2music}}.


\section{Experimental setup}
\label{sec:5}
In this section, we present the experimental setup used to evaluate the performance of our proposed Video2Music framework, including the Affective Multimodal Transformer model. We should note that, given the novelty of this task, there are no generative music systems that include video matching that we can benchmark our model to, only other music generation models. Hence, we perform extensive quantitative and qualitative evaluations to set a new benchmark in the field. In our experimental setup, we divided our dataset into three distinct subsets, namely the training set, validation set, and test set, with a distribution ratio of 8:1:1, respectively. In the rest of this section, we describe our baseline models, followed by implementation details and evaluation metrics.

\subsection{Baseline models}
\label{sec:5.1}
We implement three models as baseline architectures: 1) Transformer~\citep{vaswani2017attention}, 2) Music transformer~\citep{huang2018music}, and 3) AMT without affective matching loss. We use the same chord tokenization method for all of the models, so that they could be trained on the chord dataset. The Transformer and Music Transformer models are trained solely on chords without considering video content. We should note that the first two models do not aim to match video, but purely generate music. The third baseline model, AMT, is trained on both chords and video features but excludes the affective matching loss.

For the listening test, we use Music Transformer as a baseline to compare our AMT model too. Hence, even for the Music Transformer, we do need to transform the generated chords into playable MIDI files in a (reduced) post-processing stage: we omit the use of the regression model for estimating note density and loudness. Only the third arpeggiation pattern in Table~\ref{tab:arp} is applied to the generated chord sequences.

\subsection{Implementation}
\label{sec:5.2}
In our experiment, we use CLIP (Contrastive Language-Image Pretraining), a powerful pretrained model, as a feature extractor. We freeze the weights of the bottleneck layers of the CLIP feature extractor pretrained on the ImageNet dataset~\citep{deng2009imagenet}. 

Before the training step, we preprocessed the input music videos and resized the video frames to $224 \times 224$ pixels to match the requirements of CLIP. When it comes to input video and music features, extracted on a per-second basis, we decided to set a time limit of 300 seconds. Hence, if the length of features in the time axis stretches beyond 300s, we perform clipping to fit within this limit. Conversely, if the length of features in the time axis are shorter than 300, we pad them to meet the required length.

We use Adaptive Moment Estimation (Adam) as our optimizer when training the Transformer model, with the initial learning rate set to 1.0. We use LambdaLR Scheduler to decay the learning rate. The betas are set to $(0.9, 0.98)$. The value of \( \lambda \) has been set to 0.4 after careful evaluation on the validation set. Finally, we used the PyTorch as a deep learning framework to implement Transformer model and regression model. All experiments were performed on a workstation with NVIDIA Tesla V100 DGXS 32 GB GPUs.

\subsection{Performance Measures}
\label{sec:5.3}

\subsubsection{Metrics for Objective Evaluation}
\label{sec:5.3.1}
To measure the inference accuracy, we adopt \( Hits@k \). This metric allows us to evaluate the generated chord progressions by calculating the ratio of the reference chord presence among the top \textit{k} candidate chords predicted by the model, where \( k \) = 1, 3, and 5. In our case, the reference chord is the ground truth chord from chord progression estimated by chord transcription model on our original audio data. \( Hits@k \) is a widely used metric for evaluating rank-based methods~\citep{yin2017spatial, yin2018joint, wang2018neural} and is also employed in music generation to assess the quality of generated results~\citep{zeng2021musicbert}. \( Hits@k \) is calculated as follows:

\begin{equation}
Hits@k = \frac{1}{n} \sum\nolimits_{i=1}^{n} \mathds{1}(rank_i \leq k)
\end{equation}
where \( n \) represents the number of samples, \( \mathds{1}( \cdot) \) denotes an indicator function that returns 1 if the rank of the target is less than \( k \), and 0 otherwise.

We included the affective matching loss, as formulated in Equation~\ref{eq:loss_emotion}, as a metric to gauge the alignment between the music and video. This metric assesses how effectively the emotions elicited by the generated chords match the emotions expressed in the video.

\subsubsection{Metrics for Subjective Evaluation}
\label{sec:5.3.2}
We conducted a comprehensive Listening test in which participants were asked to rate 20 music videos in which the music was generated by our system. They were asked to rate a number of questions on a 7-point Likert scale (1 being extremely poor and 7 representing excellent). 
These questions are aimed to offer insights into the musical quality as well as the music-video alignment: 
\begin{itemize}
	\item \textit{Overall Music Quality (OMQ)}: How high is the overall quality of the generated music (independently of the video content)?
     \item \textit{Music-Video Correspondence (MVC)}: How well are the video and music matched overall?
     \item \textit{Harmonic Matching (HM)}: How well does the harmony match the video?
     \item \textit{Rhythmic Matching (RM)}: How well does the tempo match the video?
     \item \textit{Loudness Matching (LM)}: How well does loudness match the video?
     
\end{itemize}

\section{Results}
\label{sec:6}
We performed both objective and subjective experiments, using the test set of our newly proposed dataset, MuVi-Sync for the task of music (chord) generation for videos.

\subsection{Objective evaluation}
\label{sec:6.1}
For the objective evaluation, our aim was twofold: firstly, to showcase the ability of our model to match the emotion of the video, and secondly, to show that the generated music is of high quality. To achieve this, we comparing our approach to three different baseline models: Transformer~\citep{vaswani2017attention}, Music Transformer~\citep{huang2018music}, and our proposed model without the affective matching loss.

Table~\ref{tab:chordpred} shows the \( Hits@k \) scores and affective matching loss of our proposed Affective Multimodal Transformer (AMT) with and without affective matching loss, as well as the results for two baseline models: Transformer~\citep{vaswani2017attention} and Music Transformer~\citep{huang2018music}. Our proposed model, both with and without affective matching loss shows excellent performance in terms of \( Hits@k \), indicating that the generated chords of of good quality. Both AMT models outperform the baseline models. When the affective matching loss is added, the emotion predicted from the generated chords matches the emotion predicted by the video much more than any of the other models. 

To tune the hyperparameters of our proposed Affective Multimodal Transformer (AMT) model, we conducted an experiment. Table~\ref{tab:ablexp}, summarizes this experiment, whereby we systematically vary the transformer parameters, including heads and layers. Through this experimental analysis, we identified the best-performing hyperparameters on the validation set, represented in the first row of the table. The impact of increasing or decreasing the number of transformer heads and layers can be seen in the table. Each run's performance is measured using relevant metrics such as Hits@1, Hits@3, and Hits@5. By systematically altering these parameters, we successfully discerned the optimal configuration for our AMT model.

\begin{table}[ht!]
    \small
    \caption{The $Hits@k$ scores and affective matching loss of the proposed method (AMT) and baseline models on the test set.}
    \centering
    \label{tab:chordpred}
    \begin{tabular}{ccccc}
    	\toprule
    	    Model & Hits@1 & Hits@3 & Hits@5 & Affective Matching Loss  \\
    	\midrule
            Transformer~\citep{vaswani2017attention} & 0.4789 & 0.7117 & 0.8204 & 1.8366   \\
            Music Transformer~\citep{huang2018music} & 0.4965 & 0.7303 & 0.8323 & 1.8795   \\
            AMT w/o affective matching loss & \textbf{0.5142} & 0.7585 & 0.8660 & 1.6859   \\
            AMT & 0.5139 & \textbf{0.7722} & \textbf{0.8672} & \textbf{0.4662}   \\
        \bottomrule	
    \end{tabular}
\end{table}

\begin{table}[ht!] \small
    \centering
    \caption{Hyperparameter tuning experiment, reported on the validation set. The last three columns show the results for the listed parameters.}
    \label{tab:ablexp}
    \begin{tabular}{lcccccc}
        \toprule
        Configuration & 
        Transformer Heads & 
        Transformer Layers & 
        \( \lambda \) (Total Loss Weighting)  &
        Hits@1 &
        Hits@3 &
        Hits@5
        \\ 
        \midrule
        Best & 8 & 6 & 0.4 & \textbf{0.4855} & \textbf{0.7226} & \textbf{0.8381} \\ 
        \midrule
        1 & 2 & 6 & 0.4 & 0.4819 & 0.6973 & 0.8206  \\ 
        2 & 4 & 6 & 0.4 & 0.4796 & 0.7174 & 0.8269  \\ 
        \midrule
        3 & 8 & 2 & 0.4 & 0.4307 & 0.6905 & 0.8246  \\
        4 & 8 & 4 & 0.4 & 0.4603 & 0.7118 & 0.8284  \\
        5 & 8 & 8 & 0.4 & 0.4362 & 0.6401 & 0.7492  \\
        \midrule
        6 & 8 & 6 & 0.2 & 0.4261 & 0.6879 & 0.8143  \\ 
        7 & 8 & 6 & 0.6 & 0.4398 & 0.6954 & 0.8231  \\
        8 & 8 & 6 & 0.8 & 0.4485 & 0.7091 & 0.8193  \\
        \bottomrule	
        
    \end{tabular}
\end{table}

\begin{figure}[ht!] 
        \centering 
        \includegraphics[width=9cm]{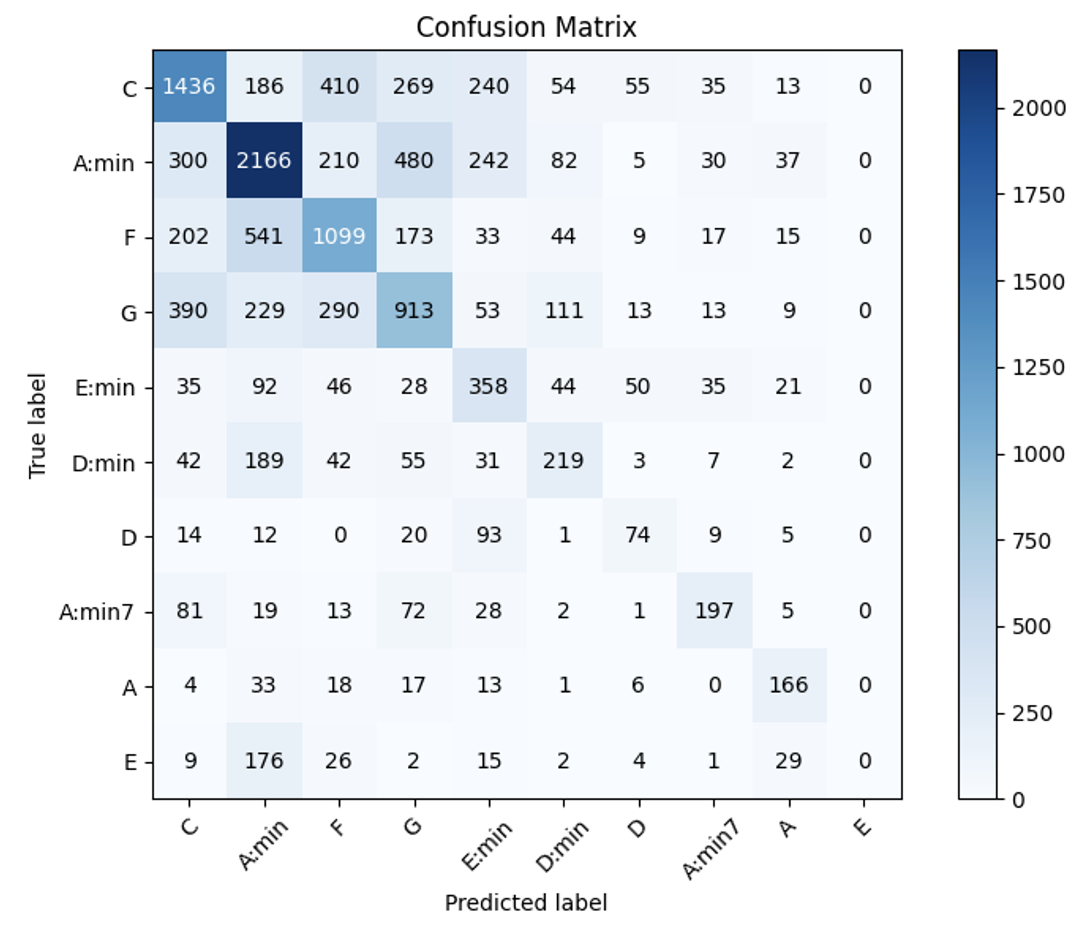} 
        \caption{Confusion matrix (chord) of our proposed model.} 
        \label{fig:confusion} 
\end{figure}  

\begin{figure}[ht!] 
        \centering 
        \includegraphics[width=9 cm]{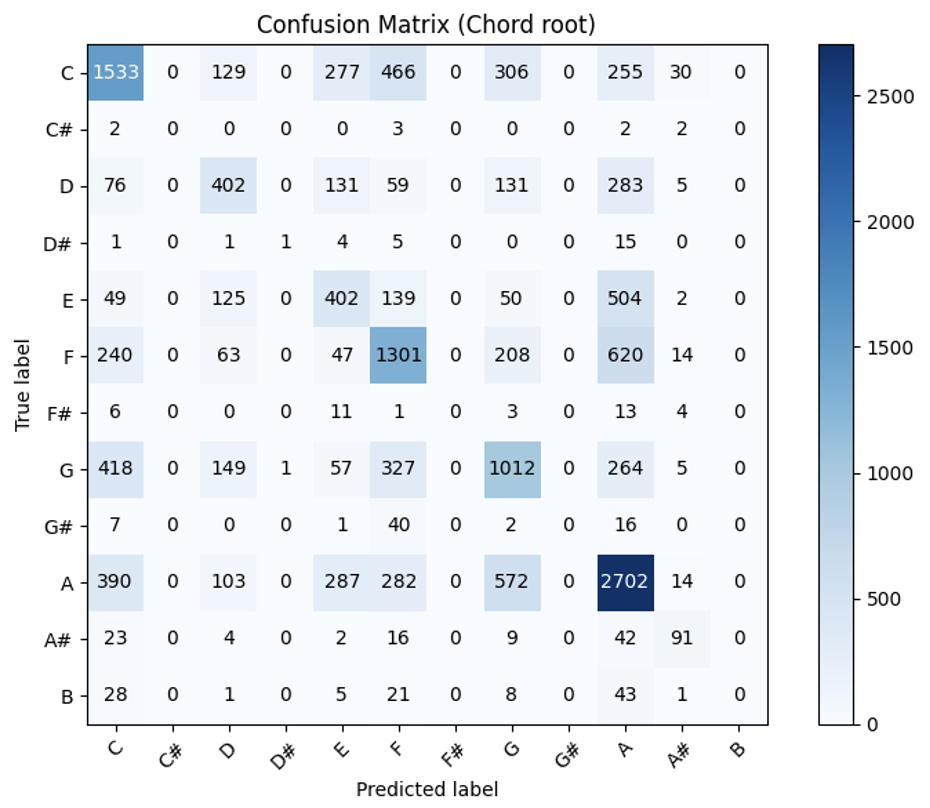} 
        \caption{Confusion matrix (chord root) of our proposed model.} 
        \label{fig:confusion_root} 
\end{figure}

We also examine confusion matrices for our model. In Figure~\ref{fig:confusion} and~\ref{fig:confusion_root}, the confusion matrices for the chord and chord root, respectively, are shown. In the former, we see a strong diagonal (correct classifications), and only a handful of mistakes. Diving deeper into the misclassifications, we see that these are musically similar chords. For instance C major and A minor both belong to the same key, and differ only in one note. The most commonly misclassified pair is G major and A minor (with 480 occurrences), this is an interesting case, all notes in the A minor chord are exactly 1 whole tone above those of the G major chord. In the key of C major, these chords are considered V and vi respectively, and often can be found in the same chord sequences, such as the popular I-V-vi-IV progression. 

\begin{figure}[ht!] 
        \centering 
        \includegraphics[width=9 cm]{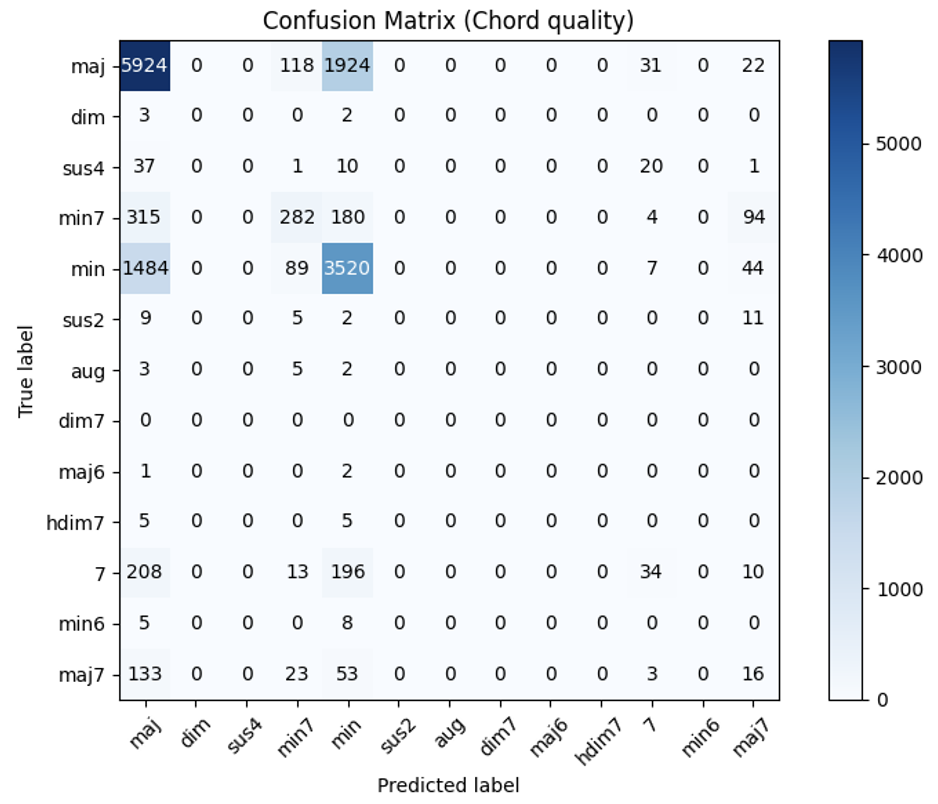} 
        \caption{Confusion matrix of the generated chord type of our proposed model matched with the chord type associated with video emotion.} 
        \label{fig:confusion_quality} 
\end{figure}

Looking at the confusion matrix for the chord root (Figure~\ref{fig:confusion_root}), we again see a strong diagonal of correct classifications. The most misclassifications occur between between perfect fifths, perfect fourths, and major/minor thirds. This again hints at the fact that our model understands music theory, as these notes often occur together in chord progressions, of which the order may be interchanged. 

The confusion matrix of our proposed model for the predicted chord type compared to the chord type that maps to the video emotion, is shown in Figure~\ref{fig:confusion_quality}. The diagonal is very present again, but there are also some closely mixed pairs, especially major versus minor. To understand this, we should explore music theory. Even in major keys (e.g. C major), it is uncommon to use only major chords. Hence, chord progression like C, G, A min, F are extremely prominent. Even though a minor chord is present in this progression, the overall sequence does not need to sound sad. Looking at some of the other pairs that are confused by the model, we see sus4 and the seventh chord type. We notice that in our mapping method (Table~\ref{tab:emochord}), these were both assigned the emotion `exciting'. Similarly, min and min7 were both assigned the emotion `sad', and maj and maj7 are both assigned `relaxing'. Even though our model may predict these different from the chord type predicted based on the emotion of the video. Both chord types still relay the same emotion. Looking at maj7 and min chord types, the first one is assigned `relaxing' and the second one `sad'. While these are separate emotions, we note that in some context they could still overlap. Overall the confusion matrices show that our model is able to match the emotion of the chords with the video. 

The above observations strongly suggests that our Affective Multimodal Transformer (AMT) excels in producing music of better quality and better matched to video content compared to existing Transformer models. We will further verify the quality of the output produced by AMT in a listening experiment.

\subsection{Listening test}
\label{sec:6.2}
We performed a listening test, where a total of 21 participants provided ratings using a 7-point Likert scale for various questions. These questions include Overall Music Quality (OMQ), Music-Video Correspondence (MVC), Harmonic Matching (HM), Rhythmic Matching (RM), and Loudness Matching (LM). To generate the final rating scores, we calculated the mean ratings given by all participants for each of these questions. The results of the subjective evaluation (listening test) are presented in Table~\ref{tab:listening}.

\begin{table}[h]
    \caption{Listening test (Music Transformer and our proposed Video2Music framework). Ratings are based on a 7-point Likert scale for the following questions: Overall Music Quality (OMQ), Music-Video Correspondence (MVC), Harmonic Matching (HM), Rhythmic Matching (RM), and Loudness Matching (LM).
    }
    \centering
	\label{tab:listening}
    \begin{tabular}{lccccc}
	\toprule
	    Model & OMQ & MVC & HM & RM & LM \\
	\midrule
        Music Transformer~\citep{huang2018music} & 3.4905 & 2.7476 & 2.6333 & 2.8476 & 3.1286 \\
        Video2Music & \textbf{4.2095} & \textbf{3.6667} & \textbf{3.4143} & \textbf{3.8714} & \textbf{3.8143}  \\
    \bottomrule	
        \end{tabular}
\end{table}

The results highlight a clear preference of music generated by Video2Music over the Music Transformer~\citep{huang2018music} across all categories. The musical quality (OMQ) of the proposed Video2Music model was rated 4.2 on average, whereas the Music Transformer model received an average score of 3.5. This was confirmed by performing a Wilcoxon Signed-Rank test, which had a $p < 0.00001$, this confirming that the rated musical quality of our proposed model is higher. In addition, the other questions were designed to test whether the music generated by Video2Music matched the video more. Indeed, it received higher scores than the baseline model in terms of Overall Correspondence, harmonic, rhythm, and loudness matching. This is confirmed by a significant $p$-value for each of these questions. The results of the listening test thus strongly support that the proposed Video2Music model can generate high quality music that matches video.

\begin{figure}[hb!] 
        \centering 
        \includegraphics[width=12 cm]{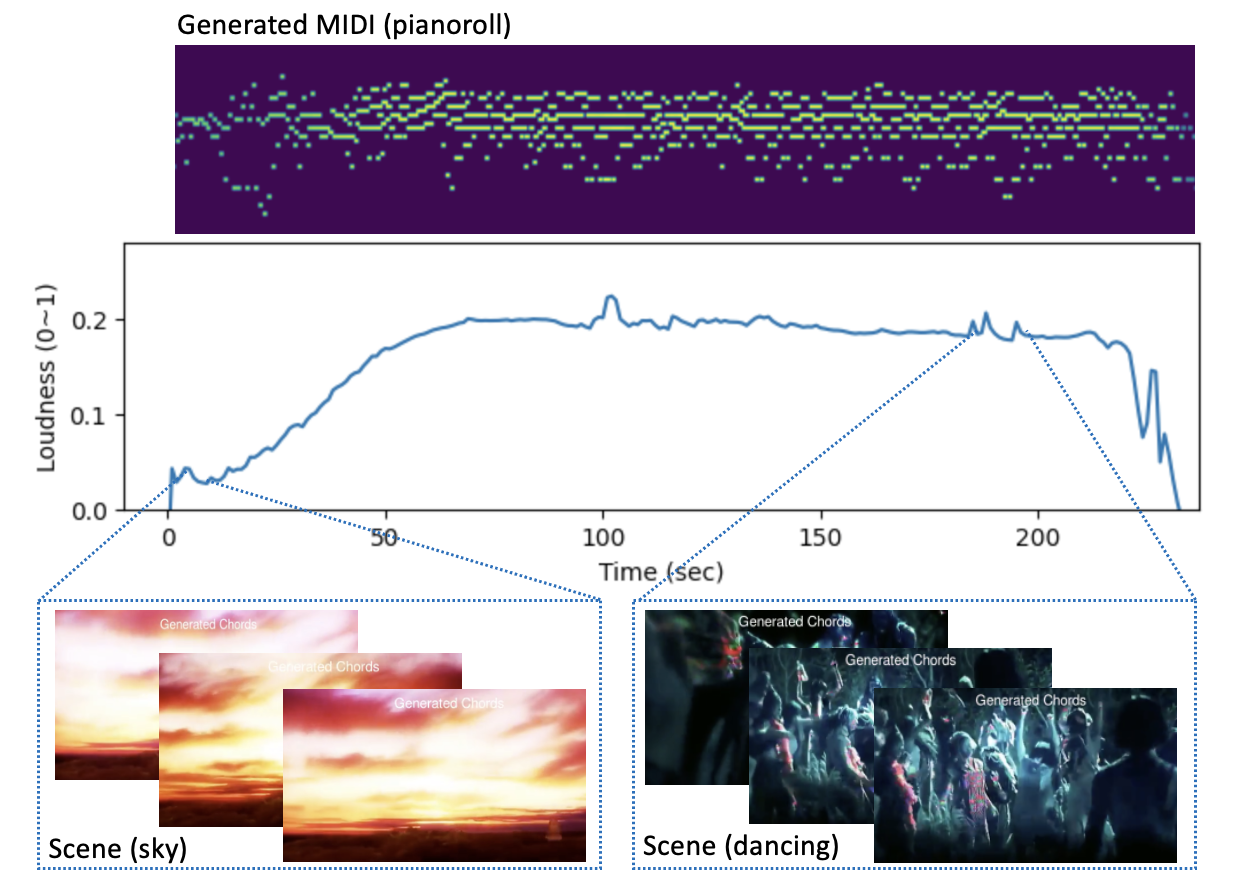} 
        \caption{Generated MIDI in pianoroll together with estimated loudness (during post-processing) for two selected video scenes (sky and dancing) from different times in the video. }
        \label{fig:gen} 
\end{figure}

Finally, Figure~\ref{fig:gen} shows a visual representation of our model's loudness and generated MIDI in pianoroll format, together with two selected video scenes (sky and dancing) from different times in the video. Notably, for the scene depicting the sky, there is a discernible pattern of low loudness and note density levels in the generated music. In contrast, during the dancing scene, the figure illustrates a distinctive increase in both loudness and note density. This demonstrates that our model's ability to dynamically adapt music to match specific video content.

\section{Conclusion}
\label{sec:7}

In conclusion, our work represents a significant stride in the field of multimodal generative systems, by introducing a novel Video2Music framework for generating music that seamlessly matches accompanying videos. This is a novel task, which will be further facilitated by our development of MuVi-Sync, a unique multimodal dataset annotated with symbolic music (transcription and chords) as well as a large array of video features, including semantic, scene offset, motion, and emotion. 

Our framework includes an Affective Multimodal Transformer (AMT) model, which fuses information from both the video modality, as well as the past generated chords, to generate the next chord in the sequence. This model includes a unique affective matching loss, and a post-processing module to adjust the music dynamically to match the video. The latter is achieved through the application of biGRU-based regression for controlling note density and loudness, based on video features, which ensures a dynamic and synchronized audio-visual experience.

Through an extensive experiment, we show that our model not only successfully generates music that aligns with the emotional tone of the video but also maintains high musical quality. These objective findings are further substantiated by a comprehensive listening study, which confirms the effectiveness of our approach in terms of musical quality and its ability to harmonize music and video content.

While our extensive experiments showcase the success of our model in generating high quality music that is emotionally aligned with the video, it is crucial to acknowledge the challenges encountered during our research. One aspect to consider is the potential impact of the accuracy of transcription algorithms on constructing the training dataset. We circumvented this issue as we opted to use chord transcription. This performs reasonably accurate, however, it limits the model to generating chords rather than polyphonic notes. This design chose was made due to the fact that polyphonic music transcription is not accurate enough and would introduce too many errors in the training dataset. This underscores the importance of future approaches that directly generate audio, not midi, e.g. such as the diffusion-based Mustango text-to-music model \citep{melechovsky2023mustango}. In future work, we aim to omnit the transcription phase and directly generate audio.

Another limitation lies in the framework's performance in cases where video content lacks clear emotional cues or exhibits complex thematic elements. The delicate balance between maintaining musical quality and accurately reflecting video emotions remains an ongoing challenge that warrants further exploration. Additionally, we recognize a specific limitation related to the perceived repetitiveness in the generated music. This arises from our current methodology, where we exclusively generate chords and arpeggiate them to produce music. To address this limitation and enhance overall musical diversity, a promising avenue for future exploration involves incorporating melody generation based on the generated chord sequences, or include more complex arpeggiation patterns and postprocessing.

It is important to consider that music often dominates the emotion perceived when watching videos, as discussed by \citet{chua2022predicting}. Purely from video cues, the model may encounter challenges in precisely determining the intended emotion, given the multifaceted nature of visual scenes. For example, a quiet, light scene may evoke both cheerful or tense emotions, depending on the context. In such instances, additional information such as scripts or lyrics could provide valuable context for a more nuanced interpretation, or we may allow the user to enter the desired emotion as input. These considerations open up exciting avenues for future research, emphasizing the need for a holistic and multimodal approach to enhance the model's understanding of complex emotional dynamics in music videos.

This work, which introduces an innovative, multimodal dimension to the field of music generation, holds great promise for various applications, including enhancing multimedia experiences, video games, as well as film and advertising videos. The MuVi-Sync dataset and the AMT model are released as open source and are available online\footnote{\url{https://github.com/AMAAI-Lab/Video2Music}}. With this work, we aim to open the door to new possibilities in the realm of music generation for videos, by offering a workable dataset, with successful baseline models. 

Looking ahead, there are several exciting directions for future exploration in this field. One potential avenue for further development is the generation of melodies based on the generated chord sequences. Incorporating melody generation to the arpeggiated chords, such as in \citep{zixun2021hierarchical}, or drums~\citep{makris2022conditional} would add another layer of musical richness and coherence to the generated compositions, enhancing the overall aesthetic quality and emotional impact of the music-video combination.
Secondly, exploring the utilization of music in the waveform presents an intriguing area for future research. By further analyzing the audio waveform itself, we can extract and leverage additional musical attributes, such as timbre, to further enhance the generated music's fidelity and expressiveness. This extension would contribute to a more comprehensive and nuanced music generation process.
Additionally, there is scope for designing and implementing a novel chord embedding method. Embedding chords into a meaningful and structured representation would facilitate the model's understanding of chord progressions and harmonic relationships. By capturing the inherent musical knowledge encoded within chord sequences, our system could generate more sophisticated and musically coherent compositions.

In summary, our Video2Music generation framework represents a significant advancement in the field, providing a powerful tool for content creators seeking to enhance their videos with personalized and seamlessly integrated background music. The future prospects of this research are promising, with opportunities to delve deeper into multi-track generation, waveform analysis, as well as innovative chord embedding techniques. By pushing the boundaries of AI-driven music generation for videos, we can continue to revolutionize the way background music is created and further enrich the audiovisual experience for both content creators and audiences.

\section*{Acknowledgements}
This project has received funding from SUTD MOE grant SKI 2021\_04\_06. 


\bibliography{paper}

\end{document}